\documentclass[article,ij4uq]{ij4uq}              %  final version
\usepackage[hang]{footmisc}
\usepackage{bbm}
\fancypagestyle{plain}{%
  \fancyhf{}
  \fancyhead[R]{\small {\it International Journal for Uncertainty Quantification}, x(x): \thepage--\pageref{LastPage} (\myyear\today)}
  \fancyfoot[R]{\small\bf\thepage }
  \fancyfoot[L]{\fottitle}
  }
\renewcommand{\dmyy}{19}
\renewcommand{\myyear}{2019}
\renewcommand{\today}{}
\begin{document}

\volume{Special Issue, \myyear\today}
\title{Multi-Fidelity modeling of Probabilistic Aerodynamic Databases for Use in Aerospace Engineering}
\titlehead{Multi-Fidelity Modeling in Aerospace Engineering}

\authorhead{J. Mukhopadhaya, B.T. Whitehead, J.F. Quindlen \& J.J. Alonso}
\corrauthor[1]{Jayant Mukhopadhaya}
\author[2]{Brian T. Whitehead}
\author[2]{John F. Quindlen}
\author[1]{Juan J. Alonso}
\corremail{jmukho@stanford.edu}
\corraddress{Stanford University, Stanford, CA, 94025}
\address[1]{Stanford University, Stanford, CA, 94025}
\address[2]{The Boeing Company, Seattle, WA}

\dataO{11/01/2019}
%\dataO{}
\dataF{}
%\dataF{}

% --------------------------------------
% Abstract
% --------------------------------------
\abstract{Explicit quantification of uncertainty in engineering simulations is being increasingly used to inform robust and reliable design practices. In the aerospace industry, computationally-feasible analyses for design optimization purposes often introduce significant uncertainties due to deficiencies in the mathematical models employed. In this paper, we discuss two recent improvements in the quantification and combination of uncertainties from multiple sources that can help generate probabilistic aerodynamic databases for use in aerospace engineering problems.  We first discuss the eigenspace perturbation methodology to estimate model-form uncertainties stemming from inadequacies in the turbulence models used in Reynolds-Averaged Navier-Stokes Computational Fluid Dynamics (RANS CFD) simulations. We then present a multi-fidelity Gaussian Process framework that can incorporate noisy observations to generate integrated surrogate models that provide mean as well as variance information for Quantities of Interest (QoIs). The process noise is varied spatially across the domain and across fidelity levels. Both these methodologies are demonstrated through their application to a full configuration aircraft example, the NASA Common Research Model (CRM) in transonic conditions. First, model-form uncertainties associated with RANS CFD simulations are estimated. Then, data from different sources is used to generate multi-fidelity probabilistic aerodynamic databases for the NASA CRM.  We discuss the transformative effect that affordable and early treatment of uncertainties can have in traditional aerospace engineering practices. The results are presented and compared to those from a Gaussian Process regression performed on a single data source. 
}


\maketitle

% --------------------------------------
% Intro/motivation
% --------------------------------------
\section{Introduction} \label{intro}

The rapid improvement of computational abilities in the recent past has increased the use of computer simulations to predict various physical phenomena. This, coupled with advances in the understanding of the underlying physics of these phenomena, has led to the development of simulations of varying complexity and computational cost that can describe the relevant QoIs at different levels of fidelity. These simulations allow for the critical assessment of engineering designs significantly earlier in the design process than what was previously possible with purely experimental design campaigns. For instance, RANS CFD simulations have now become commonplace in aerospace engineering; full aircraft configuration analyses utilizing meshes with 250-500 million cells are routinely carried out in engineering practice. The utility of these simulations is further extended by using the discrete realizations of the computational simulations to create continuous representations of the functions of interest. These continuous representations are called surrogate models and can be rapidly sampled for data-intensive methods such as uncertainty quantification (UQ) or design optimization \cite{queipo2005surrogate,gorissen2010surrogate}. Building surrogate models that can accurately represent the physical phenomena of interest can greatly reduce resources required for analyses in the design process \cite{jeong2005efficient}.

To build the most accurate surrogate models, only simulations of the highest fidelity should be used. Unfortunately, high-fidelity function evaluations are also computationally expensive and time consuming. Instead, lower-fidelity approximations are used which often don't model the physical phenomena correctly. Simulations that model different levels of complexity inject varying levels of uncertainty in their predictions \cite{peherstorfer_survey_2018}. These uncertainties, which are introduced due to inadequacies in the physical models being solved, are referred to as model-form uncertainties. Quantifying and understanding these uncertainties is essential to develop designs that can reliably meet certain performance requirements \cite{forrester_multi-fidelity_2007}. 

For a simple example, imagine trying to simulate the trajectory of a football to determine how much force is required to throw it 40 yards. In an effort to simplify the simulation, we make the assumption that the football is spherical in shape, instead of oblong. Making this simplification introduces model-form uncertainties due to the inadequacy of a sphere to capture the physics of an oblong football flying through the air. This will result in an incorrect calculation of the required force. But if the uncertainty introduced in the calculation by this simplifying assumption is known, it can be accounted for and the final force requirement can be changed (increased or decreased) to ensure that the football is thrown 40 yards with a prescribed rate of success.

Similarly, if uncertainty information is included in performance predictions, the probability that a particular design will meet certain performance requirements can be calculated. %
This is the cornerstone of Reliability Based Design (RBD) processes \cite{reliability}, which aim to replace the use of arbitrary factors of safety, with explicit quantification of probabilities of success/failure. In the context of aircraft design, the performance predictions take the form of aerodynamic databases that contain the expected forces and moments on an aircraft at all points in the flight envelope (defined by variables such as Mach number, altitude) and as a function of various parameters including aircraft orientation and the settings of multiple control surfaces. Uncertainties in these predictions can be used to create probabilistic aerodynamic databases that assign a probability distribution to each of these force and moment predictions. These probabilistic databases can then be sampled and used as inputs to a deterministic flight simulator to determine the probability that a new aircraft design will meet certain performance or certification requirements \cite{wendorff_combining_2016}. Designing to maximize the probability of success provides a more robust design optimization framework \cite{ng_multifidelity_2014,multif} and supplies additional information to the designer that can be used to make better design decisions.

During the typical aerospace design process, different kinds of performance analysis tools are used at different stages. Lower-fidelity computer simulations trade lower accuracy for faster computations and are useful at the very early stages of the design process when the geometry of the aircraft is not well defined and is subject to significant change. They are often replaced with higher-fidelity simulations as the design progresses and more details of the design are finalized. Experimental data, normally obtained through a costly wind-tunnel test, forms the most accurate representation of the phenomena analyzed and is obtained quite late in the design process. Instead of discarding the low-fidelity simulation data when higher-fidelity data is available, there exist methods to combine data from multiple fidelity levels to generate better surrogate models \cite{kennedy_predicting_2000,le_gratiet_recursive_2014}, but these methods ignore the uncertainties in the simulations that produce the data. The uncertainties they do quantify are due to prediction variance due to the surrogate model parameters. Building on \cite{huang_sequential_2006}, work has been done in using multi-fidelity Gaussian processes to incorporate Subject Matter Expert (SME) defined uncertainties, in addition to the prediction variance mentioned above, to create probabilistic aerodynamic databases \cite{wendorff_combining_2016} that can be used by traditional analysis and design methods in the aerospace industry. 

This paper addresses some of the shortcomings of these methods and provides a new framework that improves the multi-fidelity modeling techniques and eliminates the need for SME-defined uncertainties for CFD RANS calculations. Section \ref{sec:methodology} delves into the methodology used. This section first outlines the multi-fidelity Gaussian Process framework that is used to combine data from multiple data sources that have varying levels of uncertainty in their processes. Then the section explains the method used to quantify the uncertainty arising from one such data source, the Reynolds-Averaged Navier-Stokes (RANS) Computational Fluid Dynamics (CFD) simulations. Section \ref{sec:results} presents these databases for a full aircraft configuration, the NASA Common Research Model, for which data and uncertainties are generated using three separate levels of fidelity: vortex lattice, RANS CFD, and wind-tunnel experimental data. This is followed by a brief discussion of the outcomes in Section \ref{sec:conclusions} and suggestions of the way in which the methods presented and their associated uncertainties can be most effectively used in aerospace engineering moving forward.

% --------------------------------------
% Methodology
% --------------------------------------
\section{Methodology} \label{sec:methodology}

\subsection{Multi Fidelity Modeling} \label{sec:mf_modeling}

Tools of varying complexity, accuracy, and cost are often used for analysis of engineering designs. Combining data from these tools has several benefits. The cost of analysis can be lowered by performing high-fidelity simulations only in areas where the uncertainty in lower-fidelity simulations is high. Additionally, surrogate models built using multi-fidelity data require significantly fewer computationally-expensive high-fidelity evaluations to accurately model trends in the data \cite{robinson2008surrogate}. Finally, the combination of predictions obtained with analysis tools of varying levels of fidelity (in the prediction of the QoIs and their associated uncertainties) opens up the possibility for the use of multi-fidelity information fusion methods that can be significantly more accurate than any of the individual predictions alone.

\subsubsection{Gaussian Process Regression}\label{sec:GPR}
The basic building block of our multi-fidelity framework is Gaussian Process (GP) regression \cite{rasmussen_gaussian_2006}, which is a supervised learning technique used to build a surrogate model for an unknown function $y = f(\mathbf{x})$ given $n$ observed input-output pairs $\mathcal{D} = \{\mathbf{x}_i, y_i\}$ for $i \in\{1,...,n\}$. This unknown function can have multi-dimensional inputs, but must have a scalar output. These input-output pairs can be arranged in matrices $X$ and $\mathbf{y}$. If the function has an $m$ dimensional input then $X$ is an $\left (n \times m \right)$ matrix of inputs and $\mathbf{y}$ is an $\left (n \times 1 \right)$ vector of outputs.

Since these observations can be imperfect, each observation is assumed to carry some Gaussian noise associated with it such that $y_i \sim \mathcal{N}(E(f(\mathbf{x_i})),\sigma_i^2)$. Assuming that all the observations in $\mathcal{D}$ have a joint Gaussian distribution, a GP is completely defined by its mean function, $ \mu(\mathbf{x}) $, and a kernel function, $\mathbf{k}(\mathbf{x,x';\theta})$, that is parameterized by some hyperparameters $\theta$. For the purposes of this study, the squared exponential function is used
\begin{equation}
    \mathbf{k}\left (\mathbf{x,x'} \right ) = \sigma_f^2 \exp \left ( -\sum_{d=1}^{d=m}\frac{\left ( x_d - x'_d \right )^2}{2l_d} \right ),
\end{equation}
where $m$ is the dimension of the input. The hyperparameters for this kernel function are the signal variance $\sigma_f^2$ and the length scales $l_d$. The kernel function is used to create a kernel matrix $K \in \mathbb{R} ^{ n \times n}$ where $K_{ij} = \mathbf{k \left( x_i, x_j \right )}$.

To enable the GP to estimate functions with a non-zero mean, the mean of $f(\mathbf{x})$ is represented using $p$ number of fixed basis functions, $\mathbf{h(x)}$, and learned regression coefficients $\beta$. At a minimum, these basis functions include a constant term, but can have multiple polynomial terms. With these in mind, the surrogate model $Z$ evaluated at some location of interest, $\mathbf{x}_*$, can be represented as some mean value plus a zero-mean GP: 
\begin{equation}
    Z(\mathbf{x}_*) = \mathbf{h(\mathbf{x}_*)}^T\beta + \mathcal{GP}(0,K(\mathbf{x}_*,\mathbf{x}_*';\theta)).
\end{equation}
The basis functions and $n_*$ sample locations can also be arranged in matrices $X_* \in \mathbb{R} ^{ n_* \times m}$ and $H \in \mathbb{R} ^{ p \times n_*}$ such that each row of $X_*$ is a $m$-dimensional sample location and each column of $H_*$ is a $p$-dimensional result of the basis functions at the locations in $X_*$.

Combining the GP regression equations for noisy observations with those incorporating explicit basis functions, and writing in the matrix notation, a surrogate model is defined as 
\begin{equation}
    Z(X_*) \sim \mathcal{GP} (\mu(X_*), \sigma^2(X_*,X_*)),
\end{equation}
\begin{equation} \label{equ:mu_gpr}
    \mu(X_*) = H_*^T\hat{\beta} + K(X_*,X)[K(X,X)+\text{diag}(\sigma_i)]^{-1} (y-H^T\hat{\beta}), 
\end{equation}
\begin{equation} \label{equ:sig_gpr}
    \sigma^2(X_*,X_*) = K(X_*,X_*) - K(X_*,X)[K(X,X)+\text{diag}(\sigma_i)]^{-1} K(X,X_*), 
\end{equation}
where $\hat{\beta} = (H^TV^{-1}H)^{-1}H^TV^{-1}y$ is the best linear estimator for the basis coefficients and $V = K(X,X) + \text{diag}(\sigma_i)$ represents the kernel matrix at the observed points $\left ( K(X,X) \right )$ and includes the Gaussian noise that is associated with each observation $\left ( \sigma_i \right )$. The prediction from the surrogate model $Z(X_*)$ is defined by the mean $\mu(X_*)$ and the uncertainty associated with these predictions is represented by the diagonal of the $\sigma^2(X_*,X_*)$ function. To fully define the GP, the hyperparameters of the kernel function need to be learned from the data. The hyperparameters are chosen by maximising the marginal log-likeliood of the model, 

\begin{equation}
    \log~p(y|x;\theta) = -\frac{1}{2} \log|V| - \frac{1}{2}\mathbf{y^T}V^{-1}\mathbf{y} - \frac{n}{2}\log 2\pi.
\end{equation}

\subsubsection{Recursive Formulation for Multi-Fidelity Gaussian Processes}\label{sec:MF_GP}
It is often the case that simulations or experiments of a sufficiently high fidelity are too expensive to perform over the entire domain of interest for a modeled problem. In many cases, there are lower-fidelity approximations available that can be evaluated quickly to perform parameter studies. The aim of the multi-fidelity Gaussian Process is to use data from different fidelity levels to create a surrogate model that can best approximate the highest-fidelity function and its uncertainty, while reducing the required number of high-fidelity function evaluations. 

Assume there are $s$ information sources $f_t(\mathbf{x})$, where $t\in\{1,2,...,s\}$, and the function at the highest fidelity level, $f_s(\mathbf{x})$, is being approximated using a Gaussian Processes, $Z_s(\mathbf{x}) \sim \mathcal{N}(\mu_{s}(\mathbf{x}), \sigma_s^2(\mathbf{x}))$. An auto-regressive formulation of the multi-fidelity framework is used. This was first put forward in \cite{kennedy_predicting_2000} and was improved upon by \cite{gratiet_multi-fidelity_nodate} to reduce computational cost and improve predictions. The GP approximation at the $t$ fidelity level is modeled as

\begin{equation}
    Z_t(\mathbf{x}) = \rho_{t-1}(\mathbf{x})Z_{t-1}(\mathbf{x}) + \delta_t(\mathbf{x}),
\end{equation}
\begin{equation}
    \rho_{t-1}(\mathbf{x}) = g_{t-1}^T(\mathbf{x})\beta_{\rho_{t-1}},
\end{equation}
where $g_{t-1}(\mathbf{x})$ is a set of $q$ basis functions, similar to $h(\mathbf{x})$ in the previous section, $\beta_{\rho_{t-1}}$ is the learned regression coefficients, and $\delta_t(\mathbf{x})$ is modeled using a GP. A way to interpret these terms is to consider $\delta_t(\mathbf{x})$ the additive bias and $\rho_{t-1}(\mathbf{x})$ the multiplicative bias between fidelity levels $t$ and $t-1$. To account for the different fidelity levels and their corresponding data, the subscript $t$ is added to the notation introduced in Section \ref{sec:GPR}. For example, $X_t$ refers to all the input data at level $t$. Additionally, the term $\Sigma_t = \text{diag}(\sigma^2_{i,t})$ is introduced, which refers to the noise in the outputs $\mathbf{y_t}$. 

In Appendix B of \cite{gratiet_multi-fidelity_nodate}, Gratiet presents the predictive equations for the case when the design sets are not nested ($\mathcal{D}_t \notin \mathcal{D}_{t-1}$) and the data has no process noise, such that $\Sigma_t$ is a null matrix. This work extends those equations to include process noise $\Sigma_t \neq \emptyset$, which produces the following representations for the mean and covariance equations for fidelity level $t \neq 1$ as

\begin{equation}\label{equ:mu_Zt}
\begin{split}
    \mu_{t}(X_*) = & ~ \rho_{t-1} \left ( X_* \right ) \mu_{t-1} \left (X_* \right ) + H_*^T\beta_t + \\
    & \left [ \left ( \rho_{t-1} \left (X_* \right ) \rho_{t-1} \left (X_t \right )^T \right ) \odot \sigma^2_{t-1} \left(X_*,X_t \right) + K_{t} \left(X_*,X_t \right)\right] \\ 
    & \left [ \left ( \rho_{t-1} \left (X_t \right ) \rho_{t-1} \left (X_t \right )^T \right ) \odot \sigma^2_{t-1} \left(X_t,X_t \right) + V_t \right ]^{-1} \\ 
    & \left (\mathbf{y}_t - \rho_{t-1} \left (X_t \right ) \odot \mu_{t-1} \left (X_t \right) - F_t^T \beta_t \right),
\end{split}
\end{equation}
and
\begin{equation}\label{equ:sig_Zt}
\begin{split}
    \sigma^2_{t}(X, \Tilde{X}) = & ~ \left (\rho_{t-1} \left ( X \right ) \rho_{t-1} ( \Tilde{X})^T \right ) \odot \sigma^2_{t-1} (X, \Tilde{X}) + K_t(X, \Tilde{X}) - \\
    & \left [ \left ( \rho_{t-1} \left (X \right ) \rho_{t-1} \left (X_t \right )^T \right ) \odot \sigma^2_{t-1} \left(X,X_t \right) + K_{t} \left(X,X_t \right)\right] \\ 
    & \left [ \left ( \rho_{t-1} \left (X_t \right ) \rho_{t-1} \left (X_t \right )^T \right ) \odot \sigma^2_{t-1} \left(X_t,X_t \right) + V_t \right ]^{-1} \\ 
    & \left [ \left ( \rho_{t-1} \left (X_t \right ) \rho_{t-1} ( \Tilde{X} )^T \right ) \odot \sigma^2_{t-1} (X_t, \Tilde{X} ) + K_{t} ( X_t, \Tilde{X} ) \right], 
\end{split}
\end{equation}
where $(X, \Tilde{X})$ are generic input arguments, $V_t = K_{t} \left(X_t,X_t \right) + \Sigma_t $, and $\rho_{t-1} (X) = G_{t-1}(X)^T \beta_{\rho_{t-1}}$. $G_{t-1}(X)$ is a $q \times n$ matrix where each column is a $q$-dimensional result for the corresponding $m$-dimensional row of input $X$ and $\beta_{\rho_{t-1}}$ are learned regression coefficients. For the lowest fidelity level, $t=1$, the regular GP regression equations Eq (\ref{equ:mu_gpr}) -- (\ref{equ:sig_gpr}) are used. For a set of sample locations $X_*$, the mean predictions at fidelity level $t$ is given by $\mu_t(X_*)$ and the variance in the predictions is given by the diagonal of $\sigma^2_t(X_*,X_*)$. 

To fully define the GP of each fidelity level, the regression coefficients ($\beta_{\rho_{t-1}}$ and $\beta_t$) and the hyperparameters of the kernel functions of each fidelity level need to be learned from the data. The parameter estimation equations from  \cite{gratiet_multi-fidelity_nodate} are extended for noisy observations:

\begin{equation}
    \begin{bmatrix}
    \beta_t & \beta_{\rho_{t-1}}
    \end{bmatrix} = \left [ J_t^T \left ( K_t(X_t,X_t) + \Sigma_t \right )^{-1} J_t \right ]^{-1} \left [ J_t^T \left ( K_t(X_t,X_t) + \Sigma_t \right ) ^{-1} \mathbf{y_t} \right ], 
\end{equation}
with $J_1 = H_1$ and for $t > 1$, $J_t =  \begin{bmatrix} G_{t-1} \odot \left ( \mu_{t-1} \left ( X_t \right )  \mathbf{1}_{q_{t-1}} \right ) & F_t \end{bmatrix}$. $\mathbf{1}_{q_{t-1}} \in \mathbb{R} ^{q_{t-1} \times n_t} $ is a matrix of ones. The hyperparameters of the kernel functions are learned by minimizing the negative marginal log-likelihood of each fidelity level.

\begin{equation}
    \log~p(\mathbf{y}_t|X;\theta) = -\left (\frac{1}{2} \log|V_t| + \frac{1}{2} \alpha^T V_t^{-1} \alpha + \frac{n_t}{2}\log 2\pi \right),
\end{equation}
where $\alpha = \left (\mathbf{y_t} - \rho_{t-1} \beta_{\rho_{t-1}}-F_t\beta_t \right )$.

\begin{figure}
    \centering
    \begin{subfigure}[Time taken to train the multi-fidelity GP] {
        \includegraphics[trim=80 180 112 205, clip,             width=.45\textwidth]{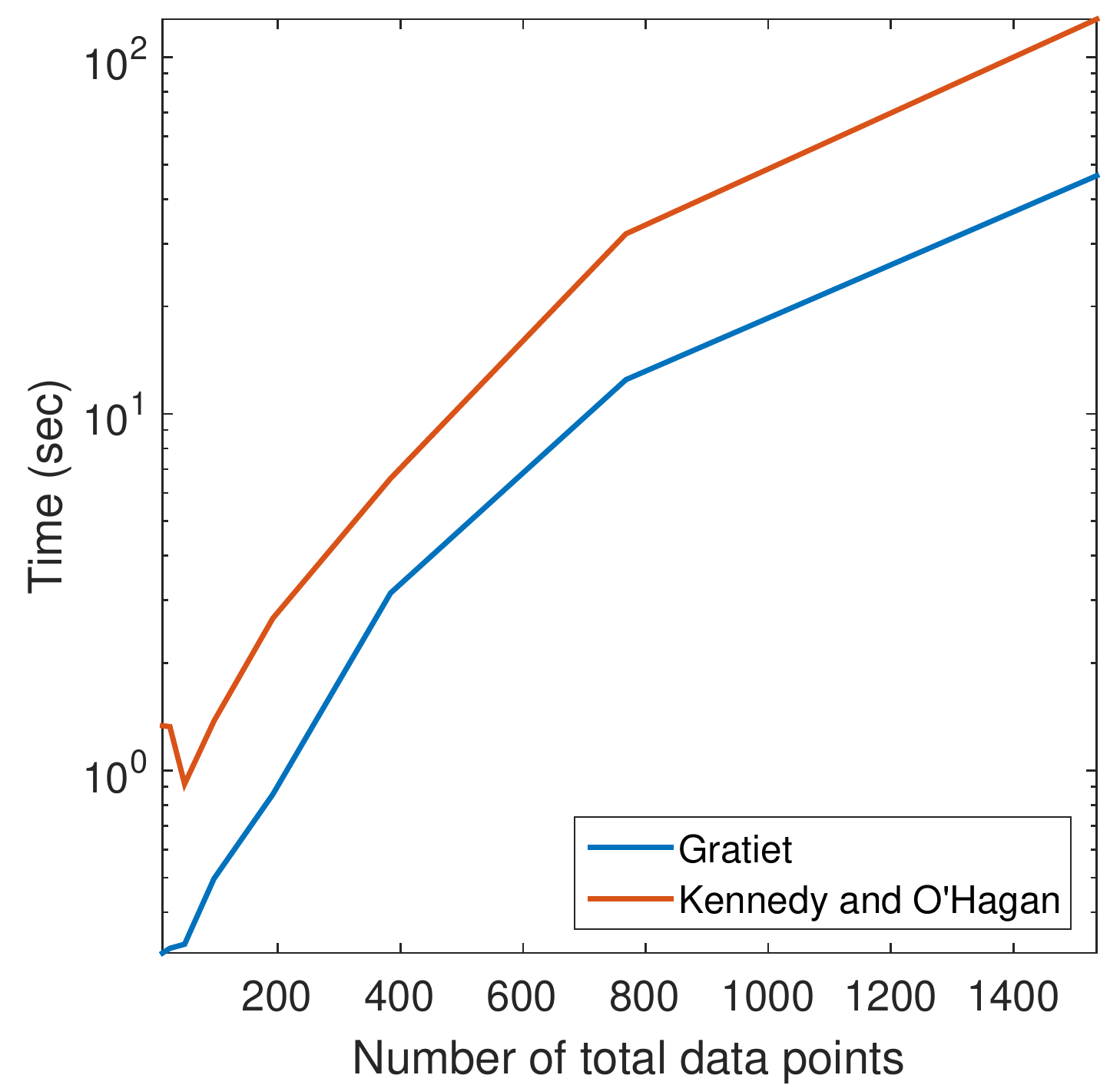} }
    \end{subfigure}
    \hfill
    \begin{subfigure}[Time taken to sample the multi-fidelity GP]{
        \includegraphics[trim=80 180 112 205, clip,             width=.45\textwidth]{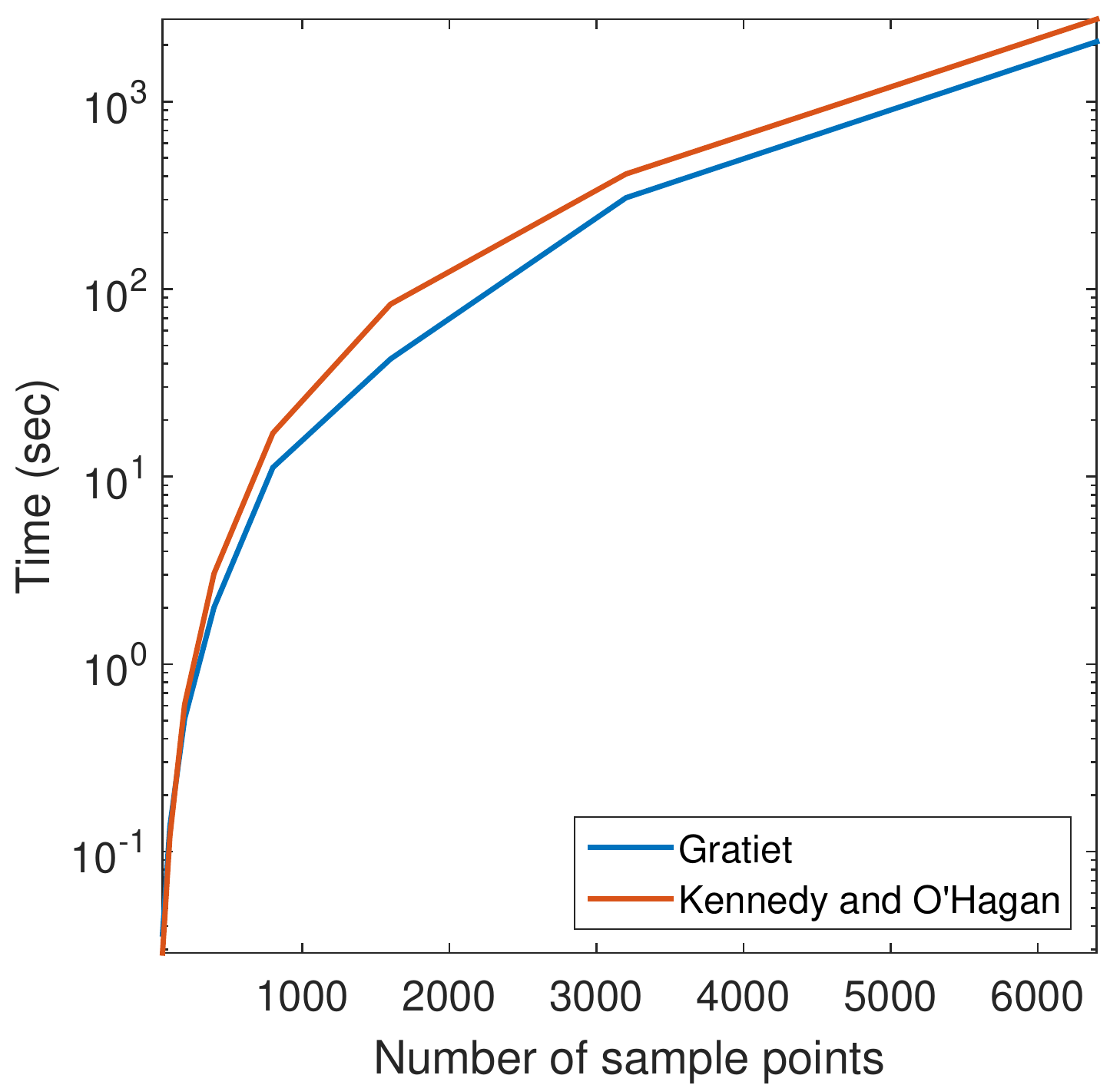} 
    }
    \end{subfigure}
    \caption{Wall clock time comparison to train and query the multi-fidelity Gaussian Process formulations put forth by Kennedy and O'Hagan \cite{kennedy_predicting_2000}, and Gratiet \cite{gratiet_multi-fidelity_nodate} \label{fig:time_comp}}
\end{figure}

As mentioned earlier in this section, the recursive formulation put forth by Gratiet \cite{le_gratiet_recursive_2014} improves on the work originally done by Kennedy and O'Hagan \cite{kennedy_predicting_2000} by reducing the computational complexity of the training and sampling steps of the multi-fidelity GP. This is achieved by splitting the dataset into each individual fidelity level instead of agglomerating the data from all levels into one set of equations. This results in having to invert smaller matrices, which greatly improves the computational cost of the process. Figure \ref{fig:time_comp} shows the comparative times for the training and sampling steps are shown. Two simple one-dimensional, analytic functions were used to generate the result: 

\begin{equation}
    f_1(x) = 0.5 \left ( 6x - 0.2\right )^2 \sin{ \left (12x -4 \right )} + 10 \left ( x - 0.5 \right ) -5 ~\text{and}
\end{equation}
\begin{equation}
        f_2(x) = 2f_1(x) - 20x + 20.
\end{equation}

In this case, the number of high-fidelity data points ($n_2$) and low-fidelity data points ($n_1$) had a constant ratio: $\frac{n_2}{n_1} = 0.2$ and the number of sample points was $5n_{1}$. These savings in computational time increase with more fidelity levels and higher dimensional functions. 

\subsection{Reynolds-Averaged Navier-Stokes (RANS) UQ Methodology overview}\label{sec:rans_uq}

Computational Fluid Dynamics (CFD) is widely used in industry to predict turbulent fluid flows of engineering interest. Turbulent flows are characterized by irregular, small-scale fluctuations in the flow variables (velocity, density, pressure) that make the fluid flow chaotic and computationally intractable to simulate exactly. While higher-fidelity approaches such as large-eddy simulations (LES) and direct numerical simulations (DNS) exist, their prohibitively large computational cost limits their adoption in industrial design workflows today. 

Instead, relatively inexpensive RANS simulations with turbulence models are most commonly used to describe turbulent flows for engineering applications. Reynolds averaging starts with the decomposition of the chaotic velocity field ($ u_i$) into its mean ($\overline{u_i}$) and fluctuating ($u_i'$) velocity components such that $u_i = \overline{u_i} + u_i'$. Here $i$ denotes a coordinate direction, $i=1, 2, 3$. This allows for the time-averaging of the Navier-Stokes equations \cite{pope_2000} which leads to a closure problem due to the resulting non-linear $\overline{u_i'u_j'}$ term that has to be modeled. This term is also known as the Reynolds stress tensor, $R_{ij}$. 

Turbulence models are concerned with predicting the behavior of the Reynolds stress tensor throughout a flow of interest (and without specific data about that particular flow) in a computationally tractable manner. To this end, these models often make simplifying assumptions about the tensor that can inject significant uncertainties into their predictions. For example, a popular assumption that is employed in numerous turbulence models is the Boussinesq approximation (also known as the linear eddy viscosity hypothesis). It assumes that $R_{ij}$ can be defined by a combination of the mean rate of strain ($S_{ij}$), an eddy viscosity ($\nu_t$), and the turbulent kinetic energy ($k$):
 
 \begin{equation}
     R_{ij} = \nu_t S_{ij} - \frac{2}{3} k \delta_{ij},
 \end{equation}
where $S_{ij} = \left ( \frac{\partial \overline{u_i}}{\partial x_j} + \frac{\partial \overline{u_j}}{\partial x_i} \right )$, $x_i$ is a coordinate direction and $\delta_{ij}$ is the Kronecker delta. This linear eddy viscosity model purports a simplified proportional relationship between the Reynolds stress tensor and the mean rate of strain tensor. It assumes that the turbulent fluid is an isotropic medium where the direction of Reynolds stresses are always aligned with the mean strain rate. %
Although reasonable for simple flows without adverse pressure gradients, this assumption can severely limit the flow features that can be predicted by the turbulence model, thus introducing uncertainty into the QoIs predicted by RANS simulations. The shortcomings of the Boussinesq assumption are particularly evident in areas of the operating envelope of an aircraft where separated flows and shock-boundary layer interactions exist. %

\begin{figure}
    \center
    \includegraphics[width=0.75\textwidth]{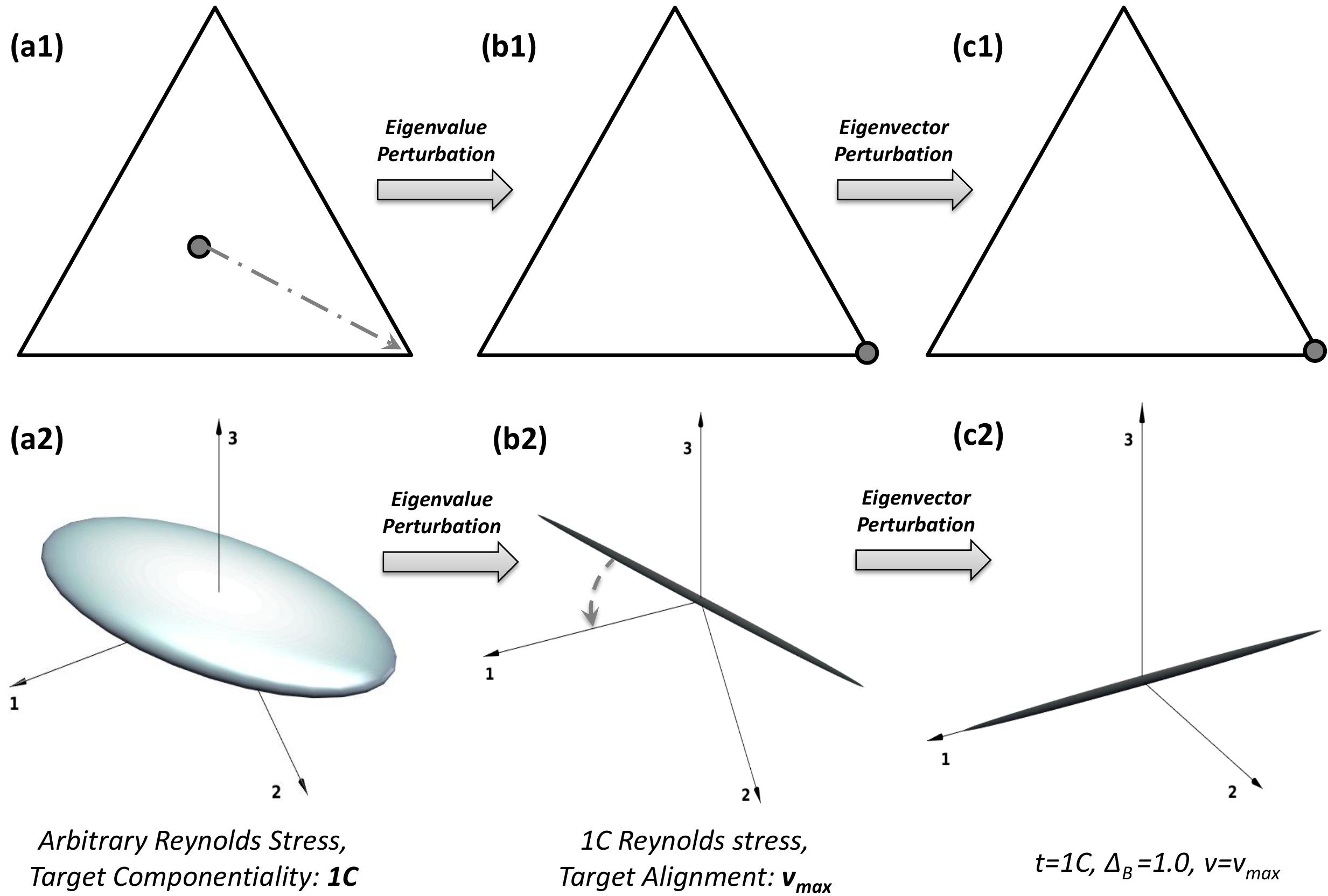}
    \caption{Schematic outline of Eigenspace perturbations from an arbitrary state of the Reynolds stress. \label{fig:pert_vis}}
\end{figure}

The eigenspace perturbation methodology was first developed by Mishra, Iaccarino, and Ghili \cite{iaccarino_eig_pert}, and was recently validated for aerospace problems of interest and implemented in the open-source SU2 solver by Mishra et al. \cite{mishra_uncertainty_2019}. It aims to quantify model-form uncertainties that arise from the use of turbulence models in RANS simulations by introducing perturbations in the eigenvalues and eigenvectors of the Reynolds stress anisotropy tensors $\left ( b_{ij} \right )$ that are predicted by the models. This methodology does not rely on any higher-fidelity data. To explain the perturbations that are introduced, the stress tensor is decomposed into its anisotropic and deviatoric components as
 
\begin{equation}
    R_{ij}=2k(b_{ij}+\frac{\delta_{ij}}{3}).
\end{equation}
Here, $k~(=\frac{R_{ii}}{2})$ is the turbulent kinetic energy and $b_{ij}$ is the Reynolds stress anisotropy tensor. The anisotropy tensor can be further decomposed into its eigenvalues and eigenvectors and represented as

\begin{equation}
b_{ij}=Q \Lambda Q^T,
\end{equation}
where $\Lambda$ is a diagonal matrix that contains the eigenvalues $\lambda_i \in \mathbb{R}$ \cite{Gerolymos2016AlgebraicPA}, and $Q$ is a matrix where the $i$-th column represents the eigenvector corresponding to $\lambda_i$. The matrices $Q$ and $\Lambda$ are ordered such that $\lambda_{1}\geq\lambda_{2}\geq\lambda_{3}$. To understand the eigenspace perturbations it helps to visualize the Reynolds stress tensor as an ellipsoid, where the axes of the ellipsoid are defined by the eigenvectors, the relative lengths along each of the axes are defined by the corresponding eigenvalues, and the size of the ellipsoid is defined by the turbulent kinetic energy. An example of such an ellipsoid, represented in a coordinate system defined by the eigenvectors of $S_{ij}$, is shown in Figure \ref{fig:pert_vis} \textbf{(a2)}. 

The perturbations are designed to exercise the limits of the physical realizability constraints placed on the Reynolds stress tensor \cite{schumann1977realizability,speziale1994realizability,2014realizability}. One way to graphically represent the constraints placed on the anisotropic part of the stress tensor is to use the eigenvalues ($\Lambda$) to project the stress tensor onto an anisotropy-invariant map, which often takes the form of a triangle. The vertices of the triangle represent the one-, two- and three-component limiting states of turbulence (referred to as the $1C$, $2C$, and $3C$ states) and all the physically realizable states of stress tensor lie within the triangle. One such representation of an anisotropy-invariant map is the barycentric map \cite{banerjee2007presentation}. The mapping allows the writing of the projection as a convex combination of the limiting states of turbulence: 

\begin{equation}
    \textbf{x} = \textbf{x}_{1C} (\lambda_1 - \lambda_2) + \textbf{x}_{2C} (2\lambda_2 - 2\lambda_3) + \textbf{x}_{3C} (3\lambda_3 + 1), 
\end{equation}
where $\textbf{x}_{1C}$, $\textbf{x}_{2C}$, and $\textbf{x}_{3C}$ represent the coordinates of the vertices that represent the one-, two-, and three-component limiting states of turbulence. For example, the stress ellipsoid shown in Figure \ref{fig:pert_vis} \textbf{(a2)} would be mapped on to the barycentric map as shown in Figure \ref{fig:pert_vis} \textbf{(a1)}.

The eigenvalue perturbations are performed such that the limiting states of turbulence are simulated. This involves perturbing the state of the tensor, sequentially, to each of the vertices of the triangle. 
The eigenvector perturbation involves changing the alignment such that the perturbed state $Q^* = Qv$, where $v$ is either:

\begin{equation}
    v_{max} = 
    \begin{bmatrix}
    1 & 0 & 0 \\
    0 & 1 & 0 \\
    0 & 0 & 1
    \end{bmatrix},~
    v_{min} = 
    \begin{bmatrix}
    0 & 0 & 1 \\
    0 & 1 & 0 \\
    1 & 0 & 0
    \end{bmatrix}.
\end{equation}

In the case shown in Figure \ref{fig:pert_vis} \textbf{(b1)}, the eigenvalues are perturbed to the 1C limiting state. This corresponds to changing the shape of the ellipsoid to be infinitely long in one direction, as shown in Figure \ref{fig:pert_vis} \textbf{(b2)}. Consequently, the eigenvector alignment is changed to maximize turbulent kinetic energy production ($v_{max}$). This is represented in Figure \ref{fig:pert_vis} \textbf{(c1)} and \textbf{(c2)}. The change in alignment of the eigenvector can not be represented in the barycentric map, which is why Figure \ref{fig:pert_vis} \textbf{(c1)} and \textbf{(b1)} are identical.

\begin{figure}
    \centering
    \begin{subfigure}[$2C$ state with $v_{max}$ eigenvector alignment] {
        \includegraphics[trim=200 40 280 30, clip,             width=000.24\textwidth]{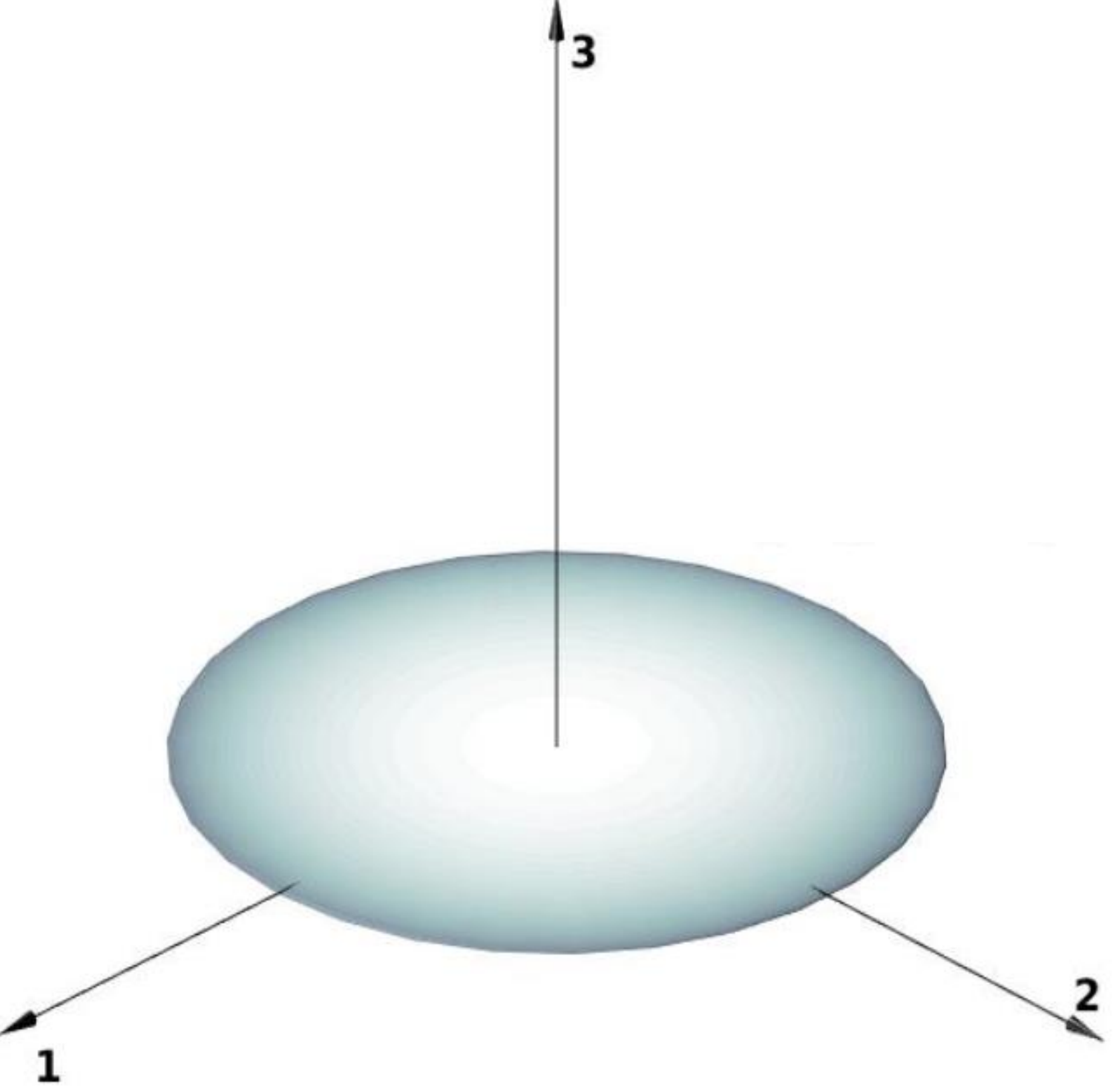}
        \label{fig:2c_max}
    }
    \end{subfigure}
    \hspace{10pt}
    \begin{subfigure}[$2C$ state with $v_{min}$ eigenvector alignment]{
        \includegraphics[trim=230 20 230 30, clip,             width=000.24\textwidth]{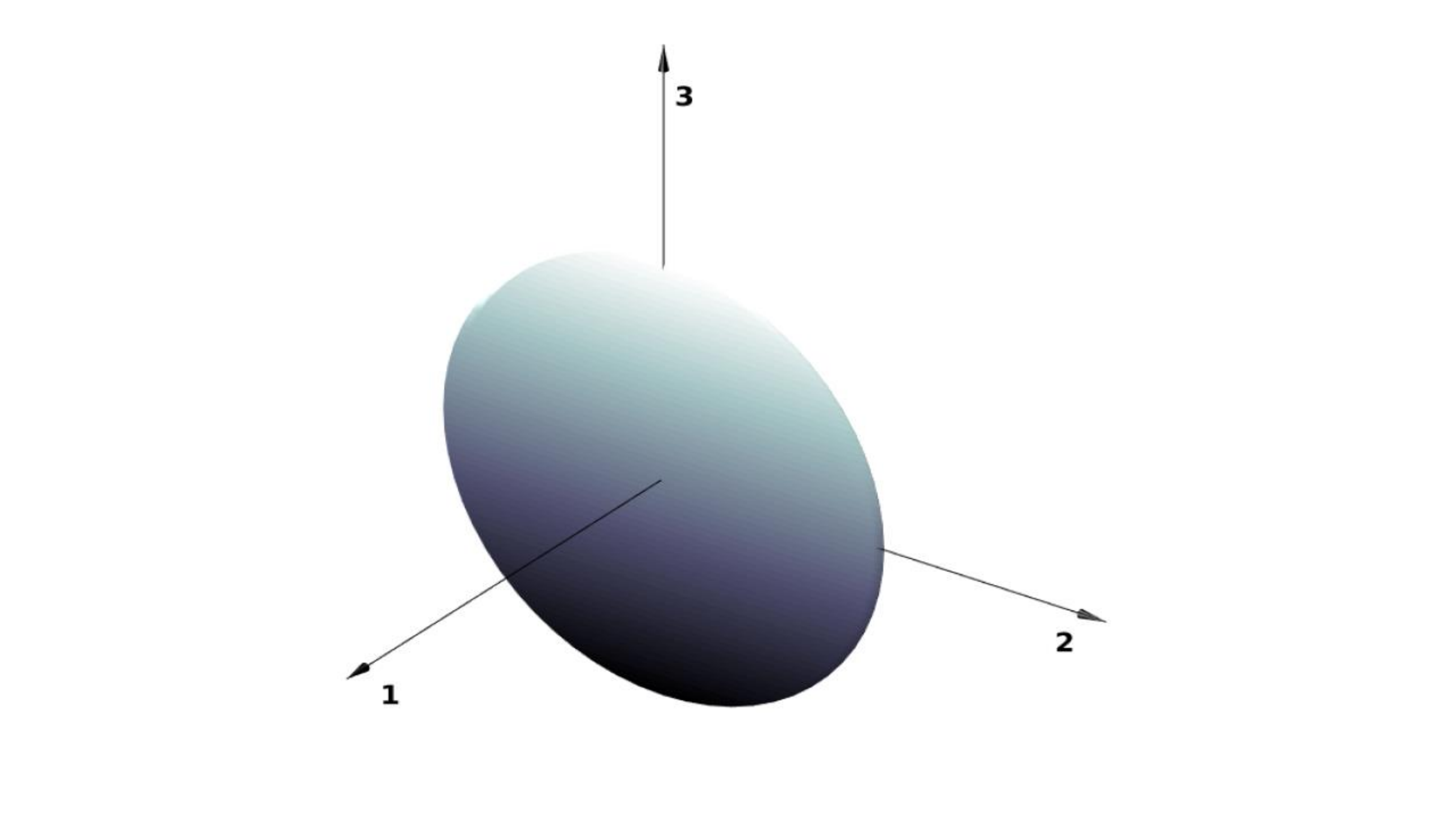} 
        \label{fig:2c_min}
    }
    \end{subfigure}
    \hspace{10pt}
    \begin{subfigure}[$3C$ isotropic turbulence state]{
        \includegraphics[trim=210 40 280 50, clip,             width=000.24\textwidth]{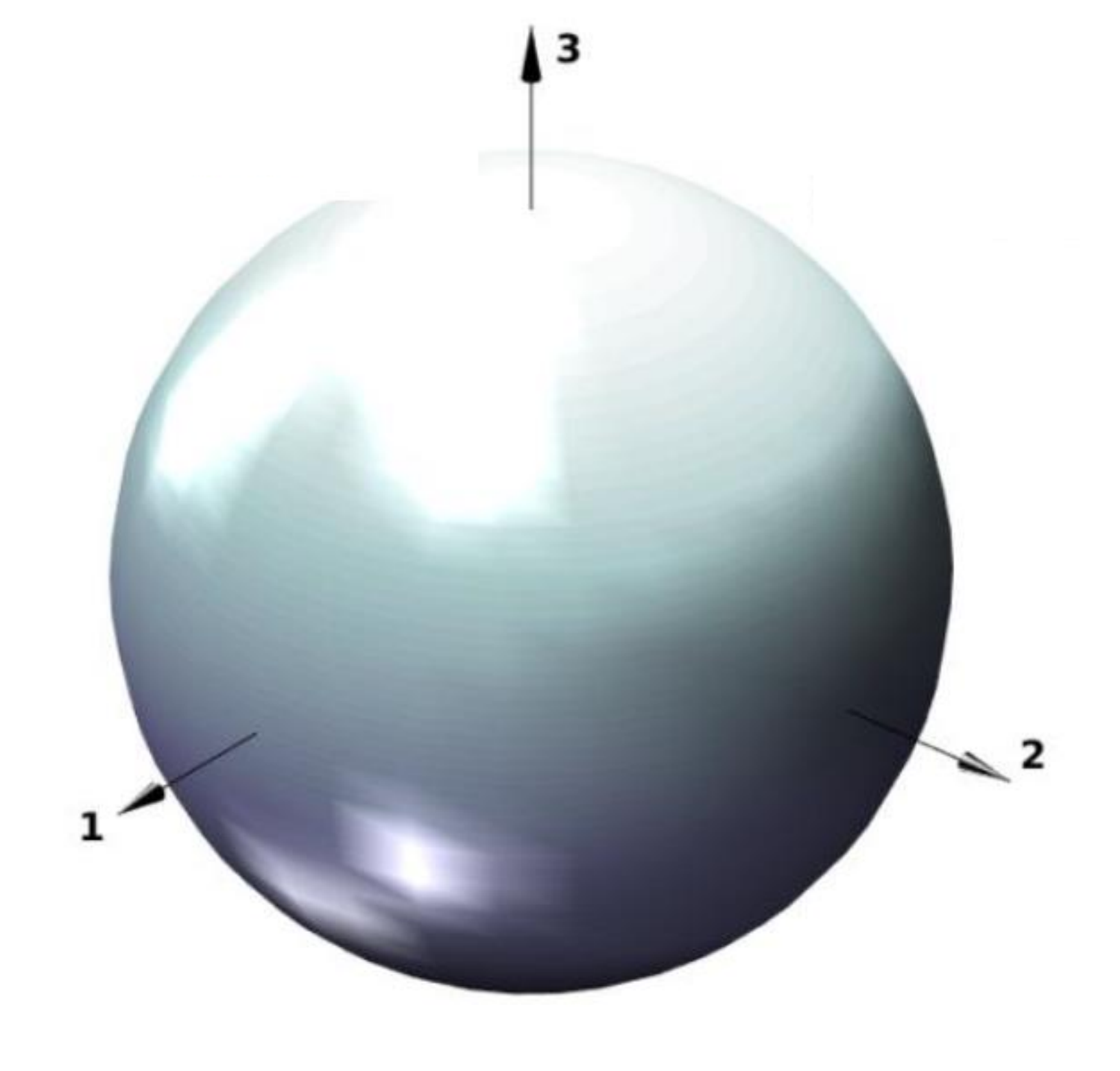} 
        \label{fig:3c}
    }
    \end{subfigure}
    
    \begin{subfigure}[$1C$ state with $v_{max}$ eigenvector alignment]{
        \includegraphics[trim=200 40 300 50, clip,             width=000.24\textwidth]{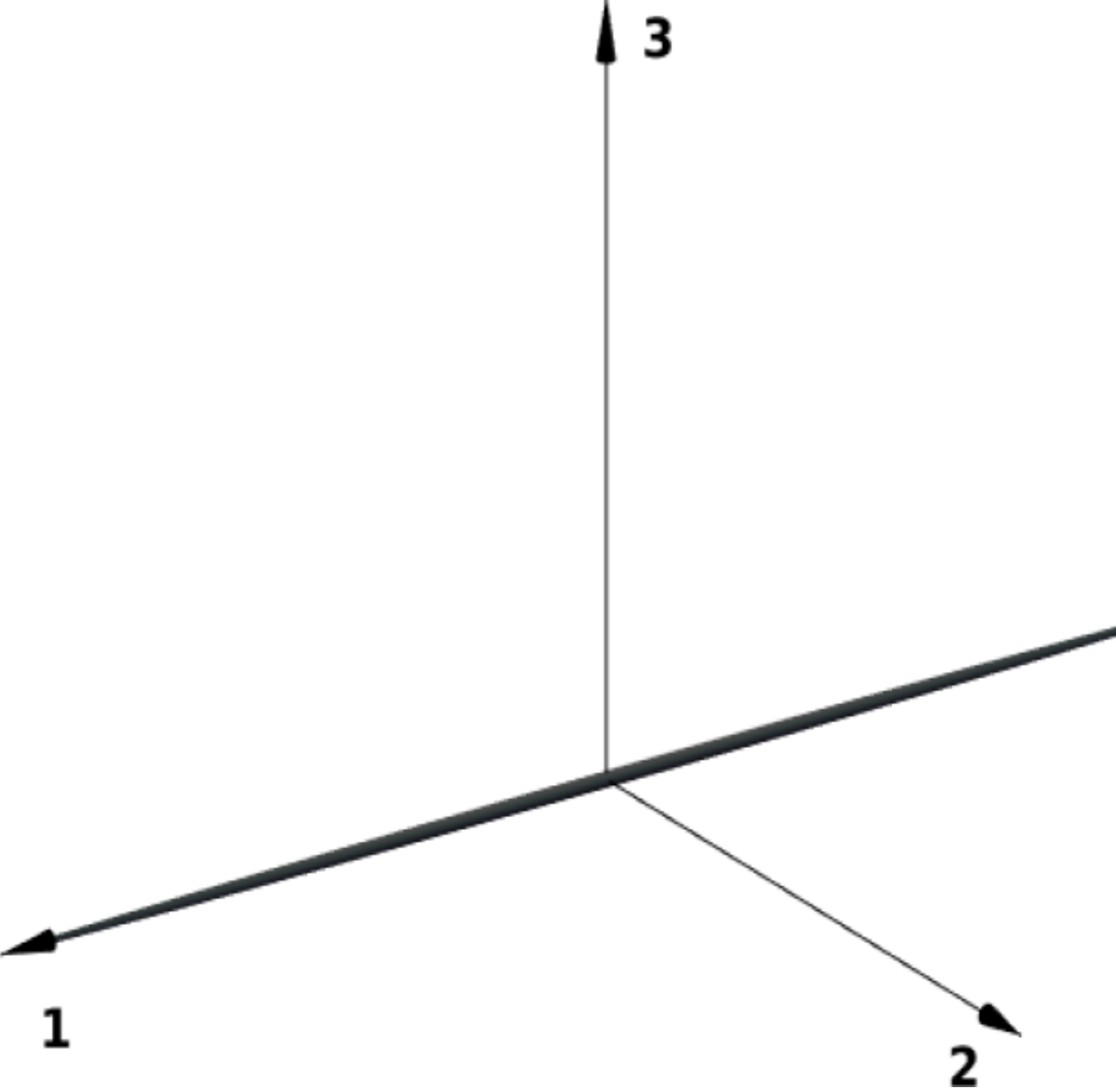}
        \label{fig:1c_max}
    }
    \end{subfigure}
    \hspace{10pt}
    \begin{subfigure}[$1C$ state with $v_{min}$ eigenvector alignment]{
        \includegraphics[trim=220 40 300 50, clip,             width=000.24\textwidth]{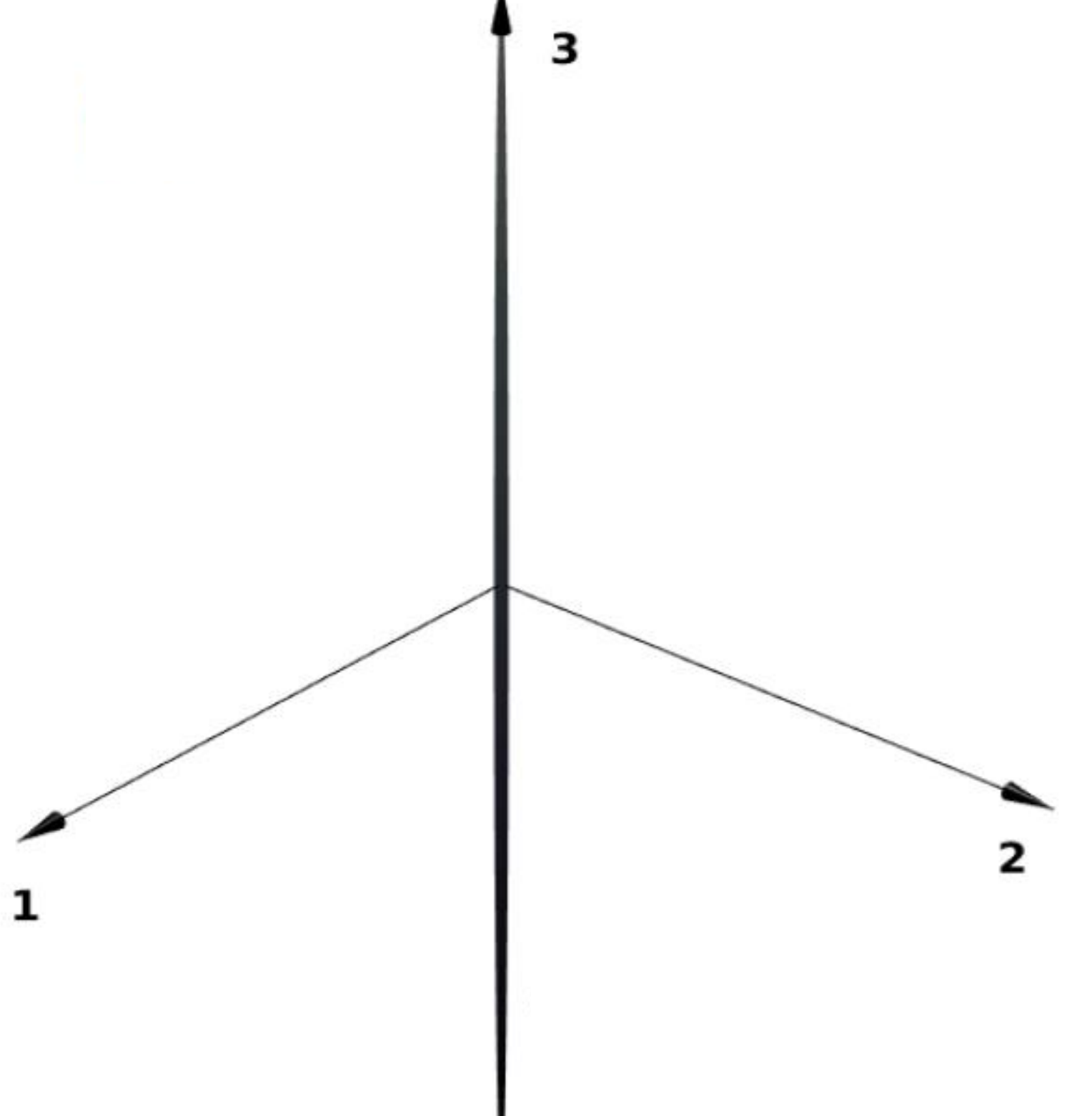}
        \label{fig:1c_min}
    }
    \end{subfigure}
    \caption{Graphical visualization of each eigenspace perturbation as Reynolds stress ellipsoids. \label{fig:all_perts}}
\end{figure}

The combinations of 3 eigenvalue perturbations (towards the $1C, 2C$, and $3C$ states) and 2 eigenvector perturbations ($v_{min}$ and $v_{max}$) result in 5 unique sets of perturbations. The resulting Reynolds stress ellipsoid shapes are shown in Figure \ref{fig:all_perts}. This number is 5, and not 6, because the $3C$ eigenvalue perturbation results in a rotationally symmetric stress ellipsoid, which means the eigenvector perturbation doesn't change anything. These 5 states mean that 5 different RANS simulations are needed, in addition to the baseline simulation with the unmodified version of the turbulence model, to provide information about the uncertainty introduced by the turbulence model. For each simulation, the same perturbation is carried out for every point in the computational domain and the solution is run to convergence. This results in 5 new realizations of the flow field in addition to the flow field predicted by the unperturbed turbulence model. The maximum and minimum values of any QoI predicted by these 6 (5 perturbed + 1 baseline) simulations create the interval bound for that QoI. These interval bounds are then used in the multi-fidelity framework as uncertainty estimates for CFD simulations. Note that the additional 5 simulations for the perturbed turbulence model can be run simultaneously and, if sufficient computational resources are available, the entire process of RANS UQ can be completed in the same wall-clock time as a typical deterministic RANS simulation.

The theoretical underpinnings of the eigenspace perturbation methodology have been discussed in detail in Mishra and Iaccarino \cite{mishra_perturbations_2019}. They prove that the eigenspace perturbations extend the isotropic eddy viscosity assumption to an anisotropic relation between the mean velocity gradients and the Reynolds stresses. This represents the most general relationship between the mean gradients and the Reynolds stresses, $ R_{ij}=\nu_{ijkl}S_{kl}+\mu_{ijkl}W_{kl}$, where $S_{kl}$ and $W_{kl}$ represent the mean rates of strain and rotation respectively. This extension enables the perturbed models to account for the effects of flow separation, secondary flows, highly anisotropic flows, etc. 

The eigenspace perturbation methodology has exhibited substantial success in a variety of engineering applications. This methodology has been successfully applied to the analysis of uncertainty in flows through scramjets \cite{emory2011characterizing}, contoured aircraft nozzles \cite{aiaajets, envelopingmodels}, and turbomachinery designs \cite{emory2016uncertainty}. The methodology has been used for the design optimization under uncertainty of turbine cascades \cite{razaaly2019optimization}. In civil engineering applications, this methodology has been applied in the design of urban canopies \cite{garcia2014quantifying,ricci2015local}. Owing to the structural similarity between the RANS and LES paradigms, this methodology has been extended and successfully applied for the uncertainty estimation of Large Eddy Simulations and scalar flux models \cite{gorle2013framework} as well.

It is important to note that this methodology provides no probability distribution information for the QoIs within these interval bounds and assuming any particular distribution would be inconsistent with the methodology. From Section \ref{sec:mf_modeling}, the multi-fidelity GP requires data to be jointly normally distributed. Consequently, for the purpose of using these interval bounds in the multi-fidelity GP framework, a Gaussian distribution of the QoIs within the bound is assumed. The QoI is given by $y \sim \mathcal{N}(\mu,\sigma^2)$ where $\mu$ is the center of the bound, and $\sigma^2$ is calculated such that $95\%$ of the interval bound predicted by the RANS UQ methodology lies at $2\sigma$ from the mean, $\mu$.

% --------------------------------------
% Numerical Implementation
% --------------------------------------
%\input{implementation}

% --------------------------------------
% Results
% --------------------------------------
\section{Applications to Probabilistic Aerodynamic Databases for Full Aircraft} \label{sec:results}

Given these building blocks, this section highlights their application to a real-world engineering problem. For this purpose, the development of a multi-fidelity probabilistic aerodynamic database for the NASA CRM aircraft is shown. It combines numerical analyses of different fidelities (vortex lattice and RANS), experimental data from wind-tunnel observations, and their associated uncertainties (provided by SMEs, the RANS UQ methodology, and experimental data uncertainties, respectively). The result is a probabilistic database that is valid within the operating envelope of the aircraft. In addition to providing mean predictions for the forces and moments that would be experienced by the aircraft under different operating conditions, this database provides variance information about the uncertainty in the predictions. 

The NASA CRM is a well-investigated full-configuration aircraft \cite{rivers_further_2012,rivers_experimental_2010} that was developed with the goal of creating a baseline geometry upon which numerous experimental and computational studies could be performed and compared \cite{morrison20094th,levy2013summary,morrison20166th,roy2017summary,tinoco2017summary}. The wealth of experimental and computational data lends itself well for the purpose of showcasing the performance of the uncertainty quantification and multi-fidelity data fusion techniques laid out in Section \ref{sec:methodology}. 

\subsection{Uncertainty Quantification in RANS CFD} \label{sec:crm_uq}
While the eigenspace perturbation methodology has been demonstrated on a variety of test cases, this is its first application on a full-configuration aircraft. The full aircraft configuration is identical (without accounting for aeroelastic deflections of the model) to the one that was tested in the wind tunnel and corresponds closely to a Boeing 777 aircraft with a fully redesigned wing. The transonic simulation conditions are described in Table \ref{NASA_CRM_test_cond}. Note that the range of angles of attack, at a free stream Mach number of 0.85, lead to non-linear physical phenomena and flow separation that can greatly increase the uncertainties in the RANS predictions. Separated flow exists for $\alpha > 4^\circ$ at this Mach number. All of the necessary RANS CFD simulations were conducted using the SU2~\cite{su2_aiaajournal} solver using the SST turbulence model~\cite{menter1994two,menter2003ten} and the previously defined perturbations to it. 

\begin{table}
\centering
    \captionsetup{justification=centering}
    \caption{Simulation conditions for the NASA CRM} 
    \begin{tabular}{|c|c|}
        \hline
        Mach Number & $0.85$ \\ \hline
        Reynolds Number & $5\times10^6$ \\ \hline
        Reynolds Length & $7.00532$ m \\ \hline
        Freestream Temperature & $310.928~\text{K}$ \\ \hline
        $\alpha$ & $-2^\circ \leq \alpha \leq 12^\circ$ \\ \hline 
    \end{tabular}
    \label{NASA_CRM_test_cond}
\end{table}

\begin{figure}
    \centering
    \begin{subfigure}[$C_L$ vs. $\alpha$] {
        \label{fig:cl_vs_alpha}
        \includegraphics[trim=80 180 112 205, clip,             width=.45\textwidth]{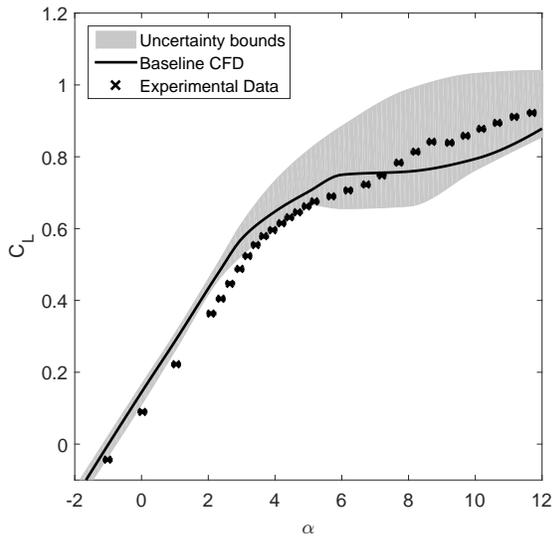} }
    \end{subfigure}
    \hfill
    \begin{subfigure}[$C_D$ vs. $\alpha$]{
        \label{fig:cd_vs_alpha}
        \includegraphics[trim=80 180 112 205, clip,             width=.45\textwidth]{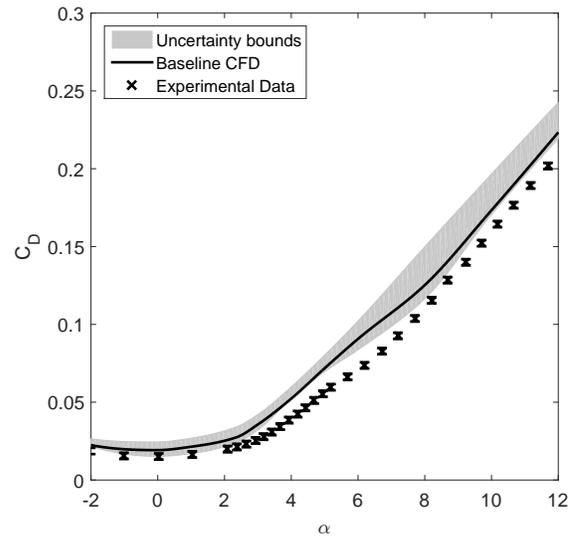} 
    }
    \end{subfigure}
    \hfill
    \begin{subfigure}[$C_m$ vs. $\alpha$]{
        \label{fig:cm_vs_alpha}
        \includegraphics[trim=75 180 112 205, clip,             width=.45\textwidth]{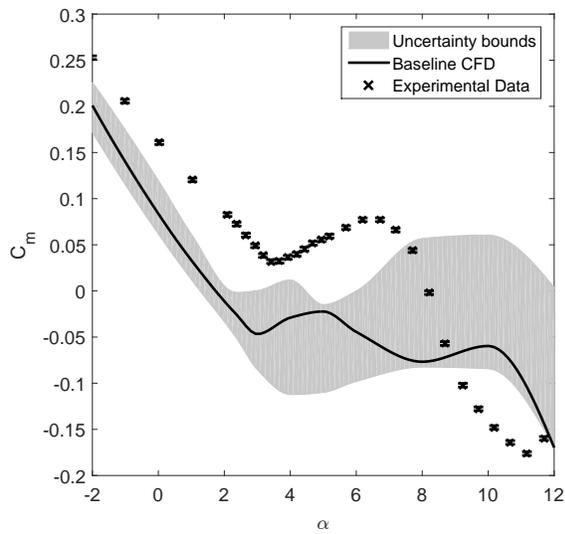} 
    }
    \end{subfigure}
    \hfill
    \begin{subfigure}[$C_L$ vs. $C_D$]{
        \label{fig:cl_vs_cd}
        \includegraphics[trim=80 180 112 205, clip,             width=.45\textwidth]{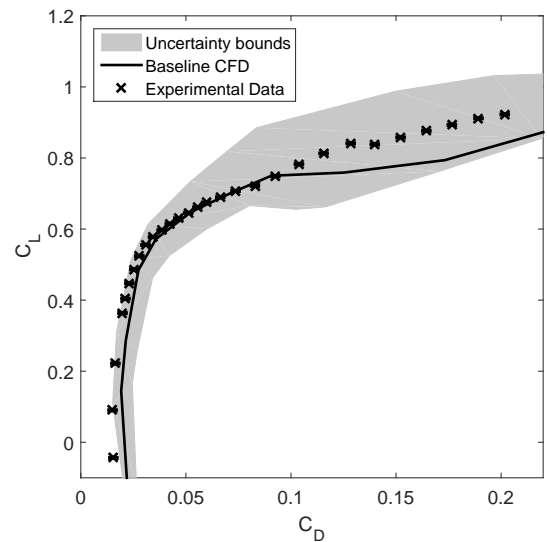} 
    }
    \end{subfigure}
    \caption{Uncertainty in force and moment coefficients as calculated by the RANS UQ methodology on the NASA CRM.\label{fig:crm_su2_uq}}
\end{figure}

In Figure \ref{fig:crm_su2_uq}, the results of these simulations are compared to wind tunnel data from the NASA Ames 11ft Wind Tunnel experiment \cite{rivers_experimental_2010}. In this figure, the solid black line represents the predictions made by the baseline SST turbulence model, the grey area represents the interval bounds predicted by the eigenspace methodology, and the black crosses represent the wind tunnel data. These wind tunnel data points have error bars associated with them but these are barely discernible on the scale of the plot. 

The performance of the UQ module is illustrated by comparing the variation of the coefficients of lift ($C_L$), drag ($C_D$), and longitudinal pitching moment ($C_m$) with respect to angle of attack ($\alpha$), as predicted by the CFD simulations to those experimentally determined. We start with $C_L$ vs. $\alpha$ in Figure \ref{fig:cl_vs_alpha}. At low angles of attack, the flow remains well attached to the aircraft body; therefore, the turbulence model does not introduce significant uncertainty in its predictions. Accordingly, the interval bounds predicted by the UQ module are relatively small. At higher angles of attack when there is flow separation over portions of the aircraft, turbulence models struggle to make accurate flow predictions due to the unsteady nature of the flow features and the structural limitations of the isotropic eddy viscosity enforced by the Boussinesq assumption. This is reflected in the growing uncertainty bounds predicted by the module. This overall trend is seen in all of the plots in Figure \ref{fig:crm_su2_uq}. 

Figure~\ref{fig:cl_vs_cd} is a typical \emph{drag polar} plot for the NASA CRM at Mach 0.85. This plot illustrates the paradigm change that RANS UQ methodologies such as the one described in this paper can bring about. A drag polar is often used in aerospace engineering in order to understand the behaviour of an aircraft across its operating range at a given Mach number. The deterministic values represented by the solid black line in the drag polar plot are used to determine the optimal operating condition for an aircraft, as well as the aircraft's safely operable range. Traditionally, only the solid black line is available to the aircraft designer and conservative factors of safety are used to build operating margins into the design. With the addition of the grey areas that represent the possible variability of the drag polar, an explicit quantification of the uncertainties can be performed. Instead of relying on a single deterministic value to design the aircraft around, the uncertainty in the performance prediction can inform the most robust optimal operating condition and the reliability of the design choices can be quantified.

Since the model-form uncertainty introduced by the turbulence model is small at low angles of attack, the CFD predictions of the force and moment coefficients should adhere more closely to the experimental data. Ideally, if there were only turbulence-model-related uncertainties in the simulations, we would see all the experimental data points lie within the grey uncertainty bounds predicted by RANS-UQ methodology. However, in the results we observe a significant deviation of the simulation data from the wind-tunnel experiments in the form of both a bias and a slight difference in the slope of the coefficient of lift curve. These differences are most prominent in the longitudinal pitching moment coefficient data in Figure \ref{fig:cm_vs_alpha}. As was discovered in \cite{levy2013summary}, the wind tunnel model of the CRM underwent significant aeroelastic deformation that affected the values recorded for the force and moment coefficients. On a swept wing like the one on the NASA CRM model, the added aeroelastic twist results in a decrease in the angle of attack of the tip region of the wing, relative to the rigid shape that was numerically analyzed. This leads to a lower $C_L$ than what was calculated. Moreover, the increase in overall twist of the wing unloads the wing-tip sections, effectively displacing the center of lift of the wing upstream, leading to the observed increased values of $C_M$ when compared to those numerically calculated. In other words, the shape of the model analyzed using numerical simulations was different than that of the real-world model. This introduced new uncertainties (of an aeroelastic nature) in the numerical predictions that were not captured in our UQ analysis. 

This discussion serves as an important reminder that regardless of the level of model (in)adequacy of a simulation method, there may be unforeseen uncertainties and errors introduced in the predictions that can cause results to deviate from real-world experiments. In this particular case, the unexpected aeroelastic deformation of the CRM model in the wind tunnel resulted in slightly different geometries being numerically and experimentally analyzed. The biases and errors introduced due to such unknowns are not explicitly derived here but they are present in the data. This means that the auto-regressive formulation of the multi-fidelity GP learns these biases and errors from the data. If the high-fidelity data is truly the most accurate representation of the physical system that is being modeled, it is good that the multi-fidelity GP can learn the bias between the lower-fidelity simulations and the high-fidelity data and compensate for it. But if there is an error in the high-fidelity data, the multi-fidelity GP will learn on the faulty data and still assume it to be the most accurate source. This brings to light the importance of the hierarchy of data sources in this formulation.

Another important point is that the interval predictions from the RANS UQ methodology only estimate the uncertainty in simulations due to the turbulence model used. The methodology cannot estimate other sources of error as it does not rely on any high-fidelity data. For example, discretization error due to insufficient mesh quality can bias the performance predictions from CFD simulations. This bias is not predicted by the RANS UQ methodology. Instead we rely on the MF GP framework to learn this and any other biases from the differences between the low- and high-fidelity data.

A key observation from Figure \ref{fig:crm_su2_uq} is that the uncertainties are not symmetric about the baseline RANS simulation (the solid black line is not in the middle of the gray area). As mentioned at the end of Section \ref{sec:rans_uq}, these bounds contain no probability distribution information. Nonetheless, for the purposes of multi-fidelity modeling, it is assumed that the distribution of the QoIs within the bounds is Gaussian and symmetric about the middle of the interval. In addition to this, the standard deviation ($\sigma$) of the Gaussian distribution is defined such that the extents of the interval bounds are $2\sigma$ away from the middle of the interval. This means that for any CFD data point with RANS UQ interval bounds, the middle of the predicted interval bound is regarded as the mean of the Gaussian distribution of the prediction, and the extent of the interval bounds are $\pm 2\sigma$ away from the mean.

\begin{figure}
    \centering
    \begin{subfigure}[$\alpha = 1^\circ$] {
        \label{fig:01aoa}
        \includegraphics[trim=50 320 190 320, clip, width=.45\textwidth]{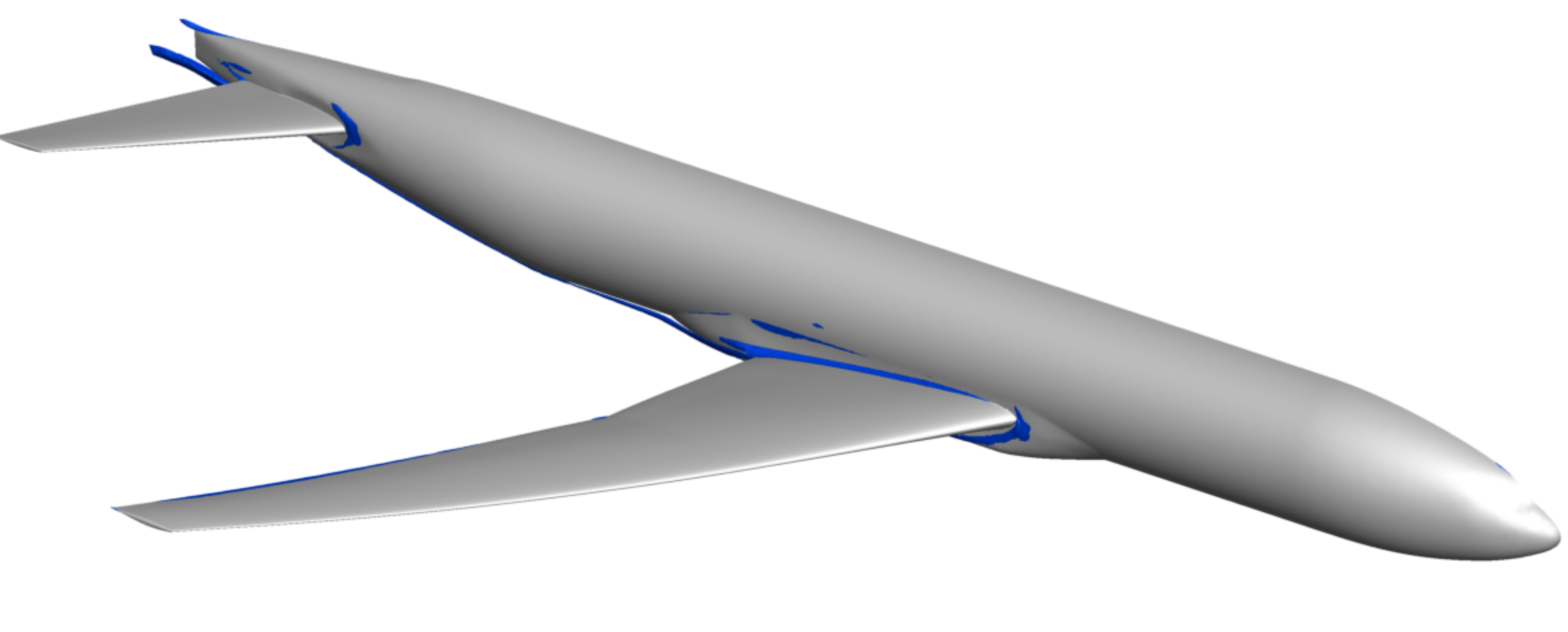} }
    \end{subfigure} 
    \hfill
    \begin{subfigure}[$\alpha = 2.35^\circ$]{
        \label{fig:02aoa}
        \includegraphics[trim=50 320 190 320, clip, width=.45\textwidth]{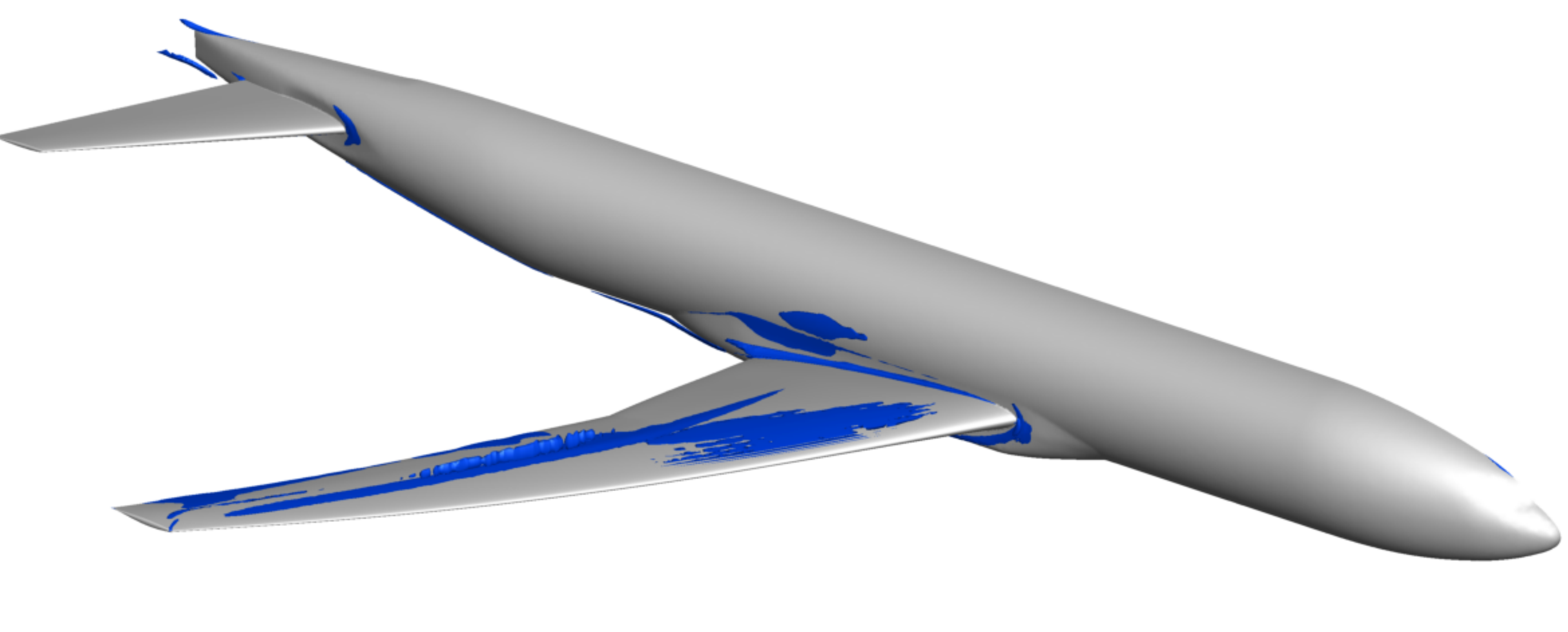} 
    }
    \end{subfigure}
    \hfill
    \begin{subfigure}[$\alpha = 3^\circ$]{
        \label{fig:03aoa}
        \includegraphics[trim=50 320 190 320, clip, width=.45\textwidth]{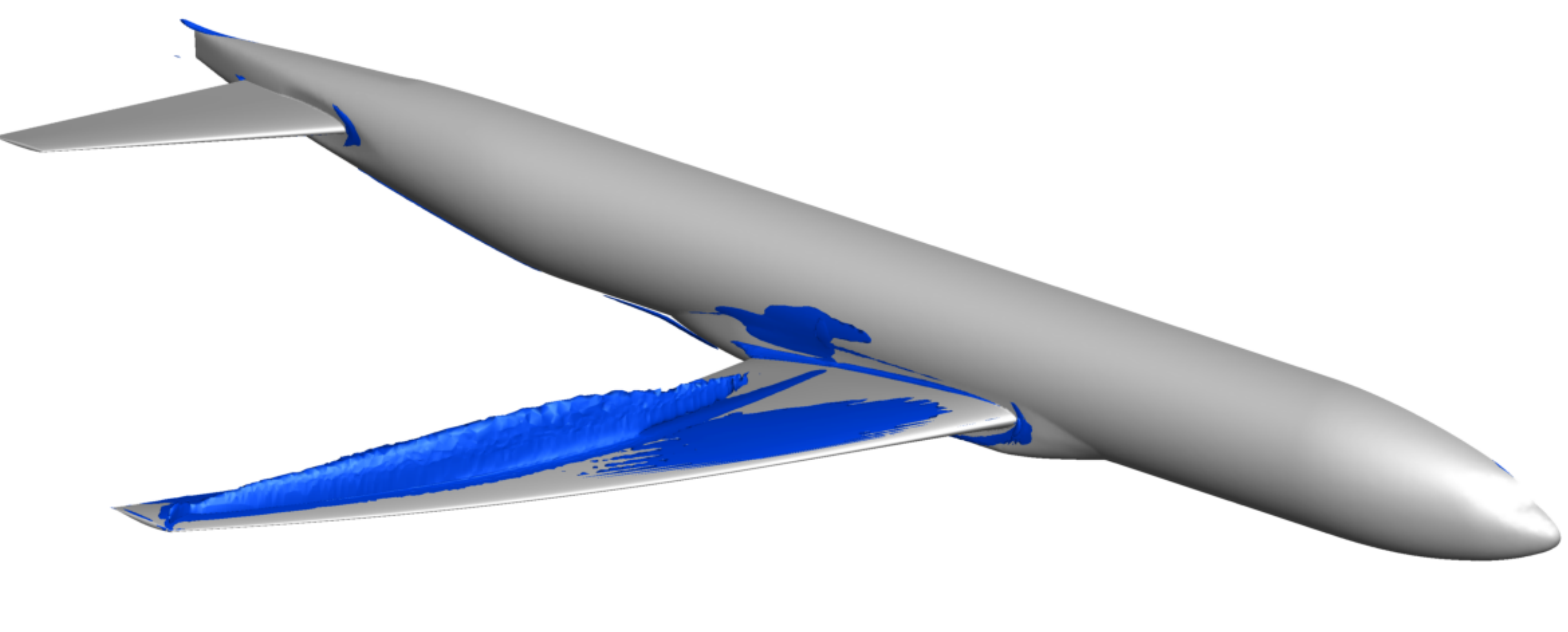} 
    }
    \end{subfigure}
    \hfill
    \begin{subfigure}[$\alpha = 4^\circ$]{
        \label{fig:04aoa}
        \includegraphics[trim=50 320 190 320, clip, width=.45\textwidth]{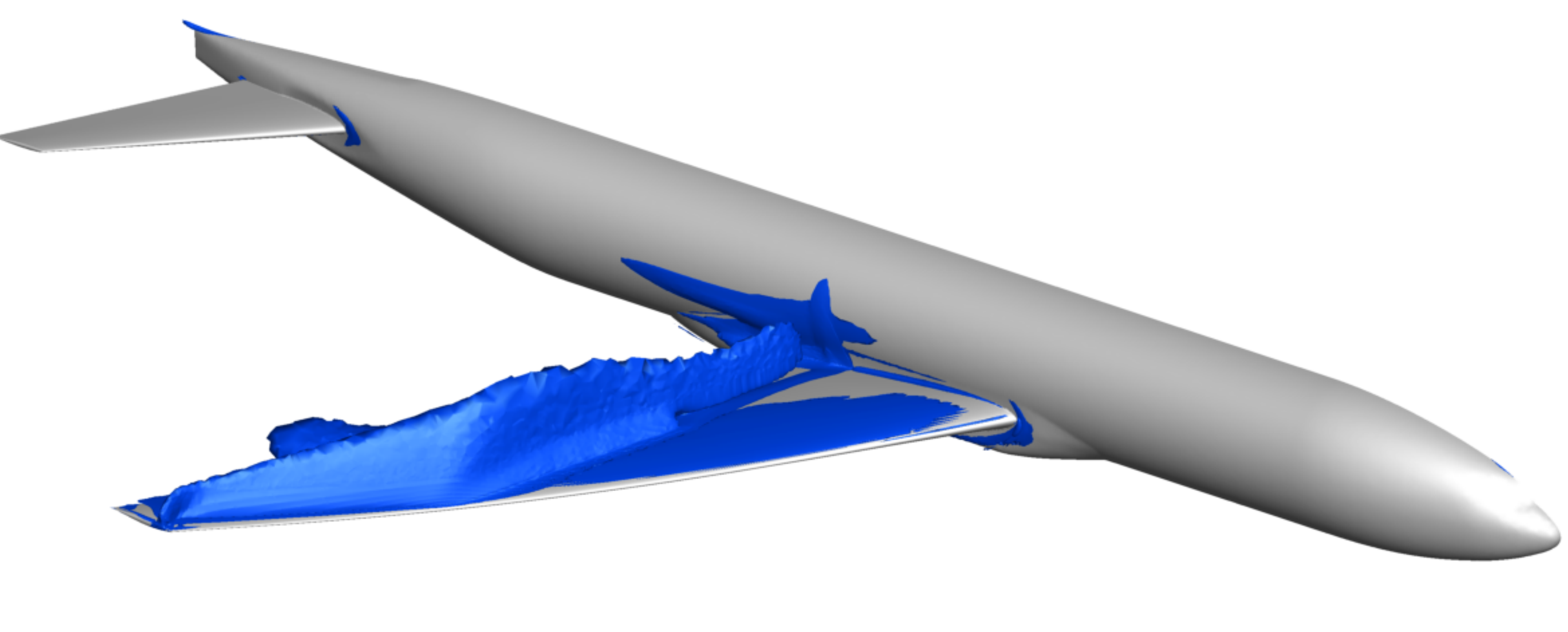} 
    }
    \end{subfigure}
    \hfill
        \includegraphics[trim=50 200 190 690, clip, width=.5\textwidth]{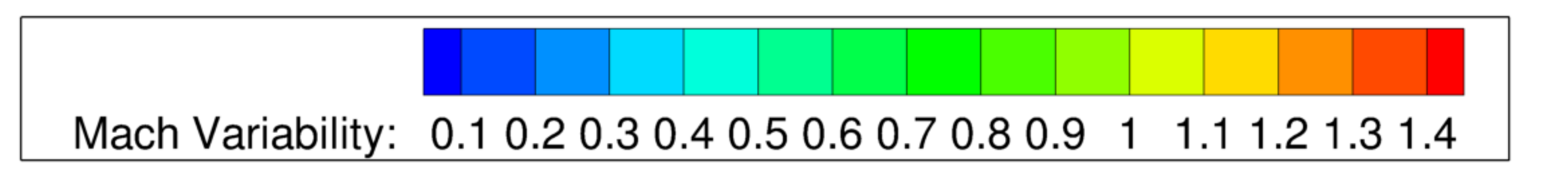} 
    \caption{Regions of high variability in the local Mach number across the perturbed simulations at various angles of attack. \label{fig:mach_isosurface}}
\end{figure}

The RANS UQ methodology doesn't only provide interval estimates on integrated quantities like $C_L$, $C_D$, and $C_m$. Since the eigenspace perturbations result in different realizations of the full flow field, the data can be post-processed to provide valuable insight into the mechanics of the turbulence model and the regions of the flow field that contribute to the resulting uncertainties. Figure \ref{fig:mach_isosurface} depicts iso-surfaces of areas where the local Mach number varies by greater than $0.2$ across all the perturbed simulations. This Mach variability $(M_v)$ is defined at every point in the computational domain as $M_v = max(M_i) - min(M_i)$ where $i$ refers to each realization of the flow field ($5$ perturbed + $1$ baseline flow fields) and $M_i$ represents the Mach number at each point in that flow field. 

At low angles of attack, Figure \ref{fig:01aoa}, the Mach variability is low and limited to small regions in the flow field. This indicates that the eigenspace perturbations do not cause major changes in the flow, resulting in smaller uncertainty bounds. As the angle of attack increases, as shown in Figures \ref{fig:02aoa} and \ref{fig:03aoa}, larger areas of variability appear where the shock would be expected, at the upper surface of the wing and away from the leading edge. This denotes an uncertainty in the shock location. This area grows rapidly until it reaches the leading edge in Figure \ref{fig:04aoa}, signalling large uncertainty bounds and reduced confidence in the CFD predictions. Such visualizations allow us to analyse the relationship between the dominant flow features and the uncertainty that they introduce in the turbulence models.

Similarly, the variability in any other flow quantity can be analyzed. Knowing the areas that contribute to the uncertainty in performance predictions can aid design decisions. These results can inform sensor placement when moving to experimental campaigns. For example, a higher density of pressure sensors can be used in areas with large coefficient of pressure ($C_P$) variability. Flow visualization techniques can be focused on areas with large velocity variability. From the perspective of turbulence modeling, these variability visualizations can shed light on the types of flow features that are hard to predict with turbulence models. This additional data processing provides more qualitative applications for the RANS UQ methodology. 

\subsection{Fusion of multi-fidelity data} \label{sec:mf_fusion}

To provide more quantitative applications of the RANS UQ methodology, we demonstrate its inclusion in the multi-fidelity GP framework. To showcase the predictive capabilities of the multi-fidelity GP framework, the CFD data is augmented with both: low-fidelity simulations, using the Athena Vortex Lattice (AVL) code \cite{drela2008athena}, and some high-fidelity experimental data, from wind tunnel campaigns \cite{rivers_further_2012,rivers_experimental_2010} used in the preceding section. For the vortex-lattice simulations, the uncertainty information is provided by subject matter experts (industry users and academics), while for the experimental data, the uncertainty intervals used are those described in the wind-tunnel campaign reports. 

\begin{figure}
    \centering
    \begin{subfigure}[Single fidelity fit] {
        \includegraphics[trim=80 180 112 205, clip,             width=.31\textwidth]{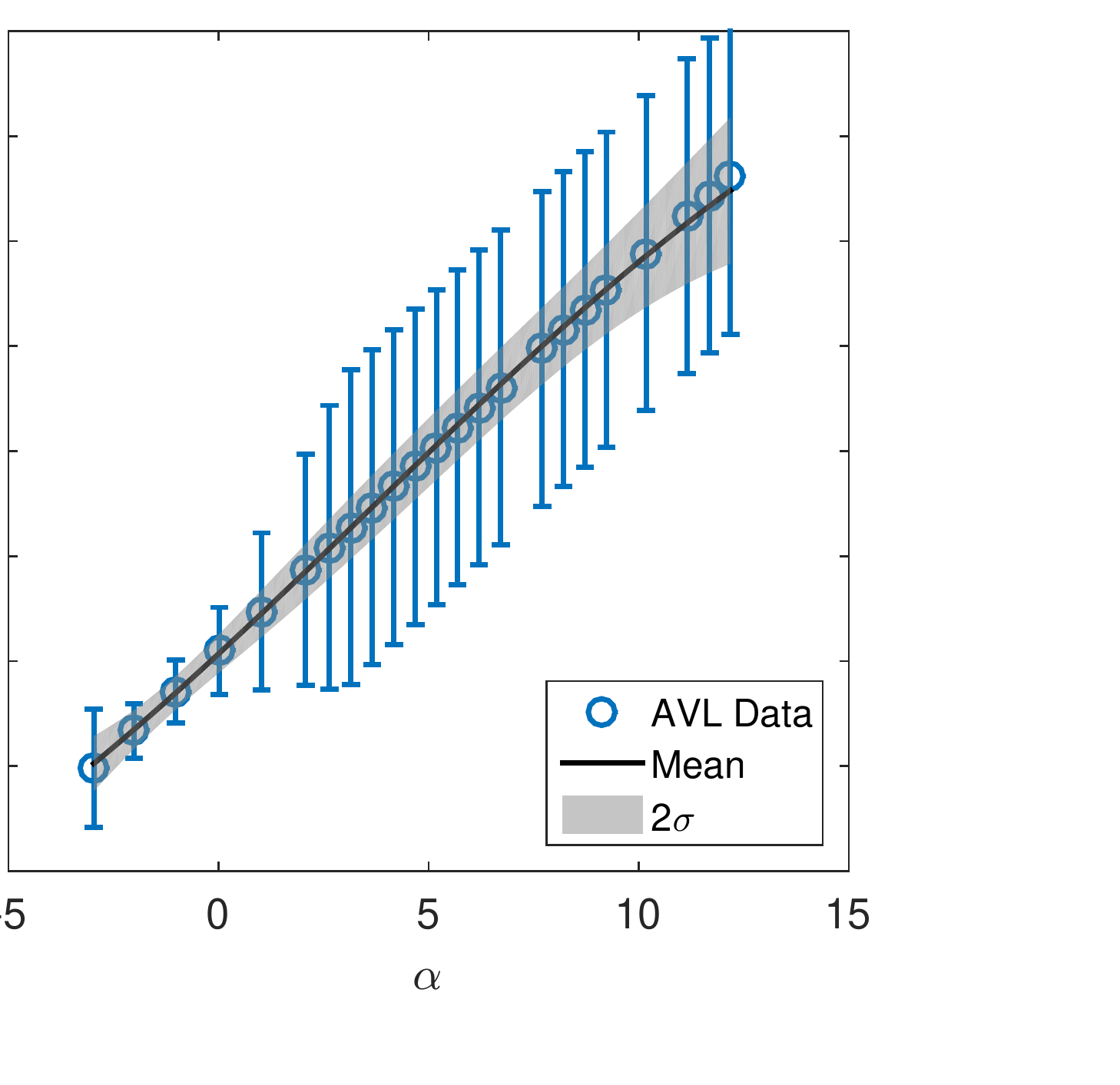} }
    \end{subfigure}
    \hfill
    \begin{subfigure}[2-fidelity fit]{
        \includegraphics[trim=80 180 112 205, clip,             width=.31\textwidth]{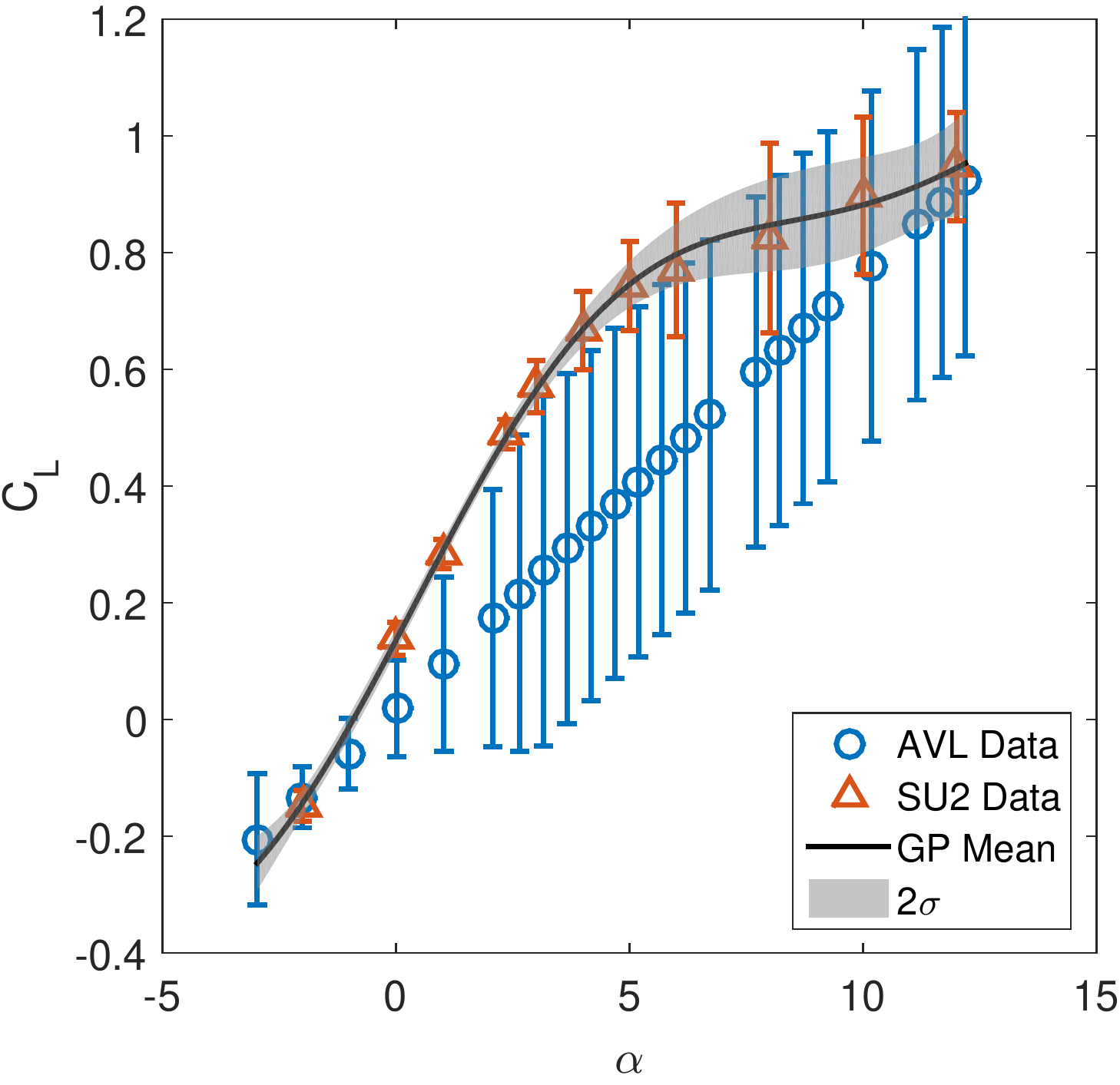} 
    }
    \end{subfigure}
    \hfill
    \begin{subfigure}[3-fidelity fit]{
        \includegraphics[trim=80 180 112 205, clip,             width=.31\textwidth]{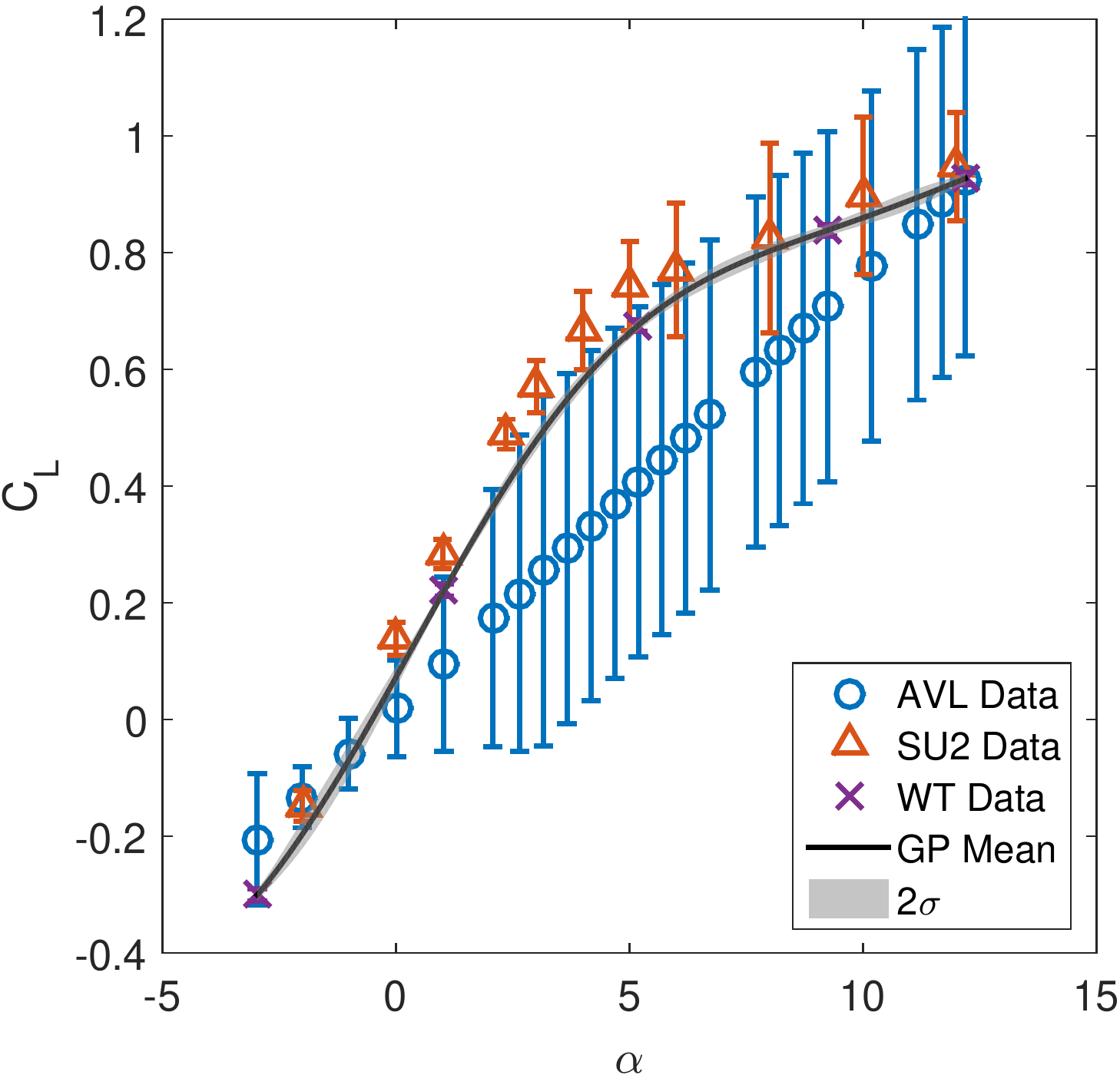} 
    }
    \end{subfigure}
    \caption{$C_L$ vs $\alpha$ for the NASA CRM, using data from multiple sources of varying fidelity.\label{fig:cl_alpha_mf}}
\end{figure}

\begin{figure}
    \centering
    \begin{subfigure}[Single fidelity fit] {
        \includegraphics[trim=80 180 112 205, clip,             width=.31\textwidth]{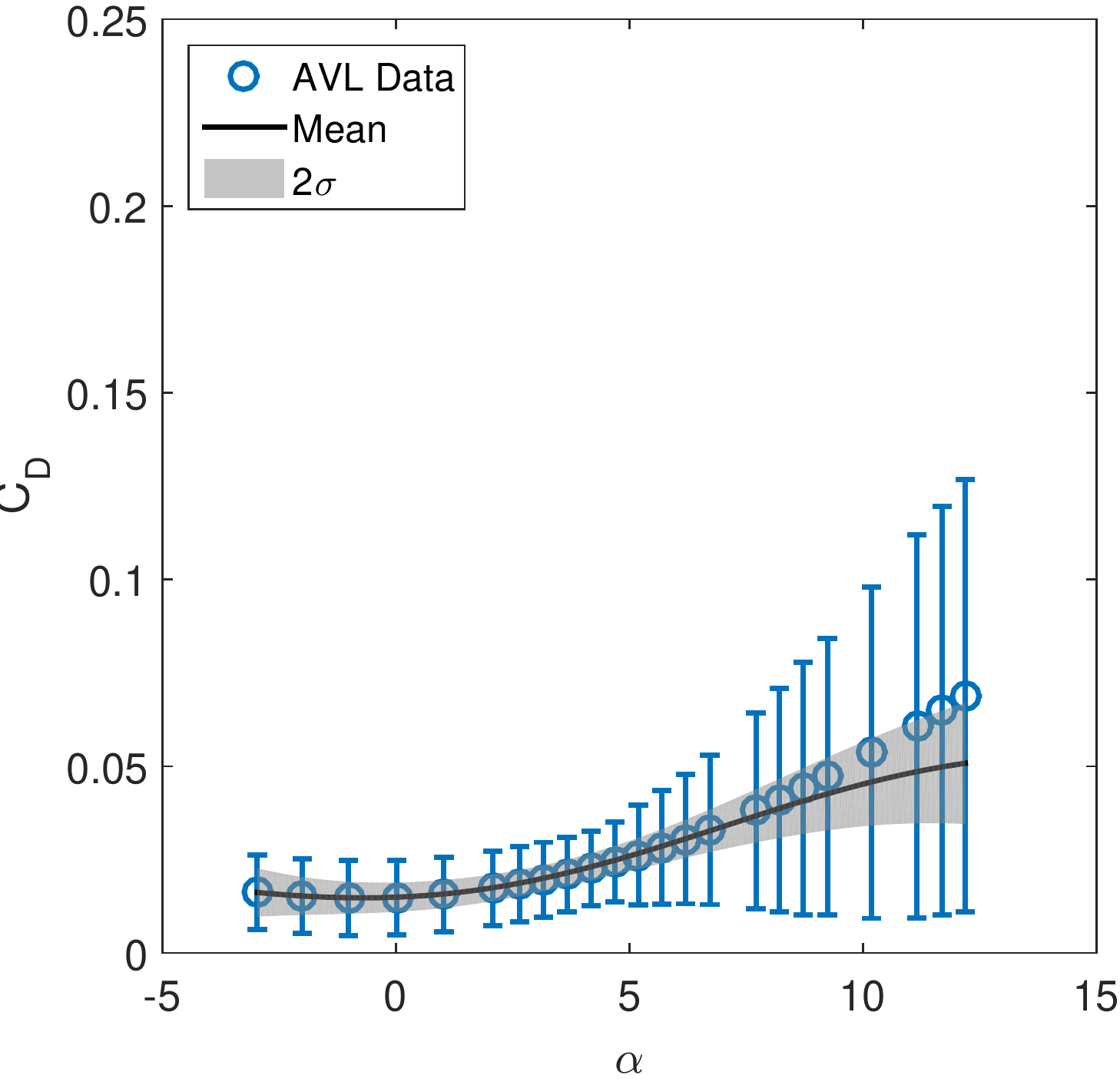} }
    \end{subfigure}
    \hfill
    \begin{subfigure}[2-fidelity fit]{
        \includegraphics[trim=80 180 112 205, clip,             width=.31\textwidth]{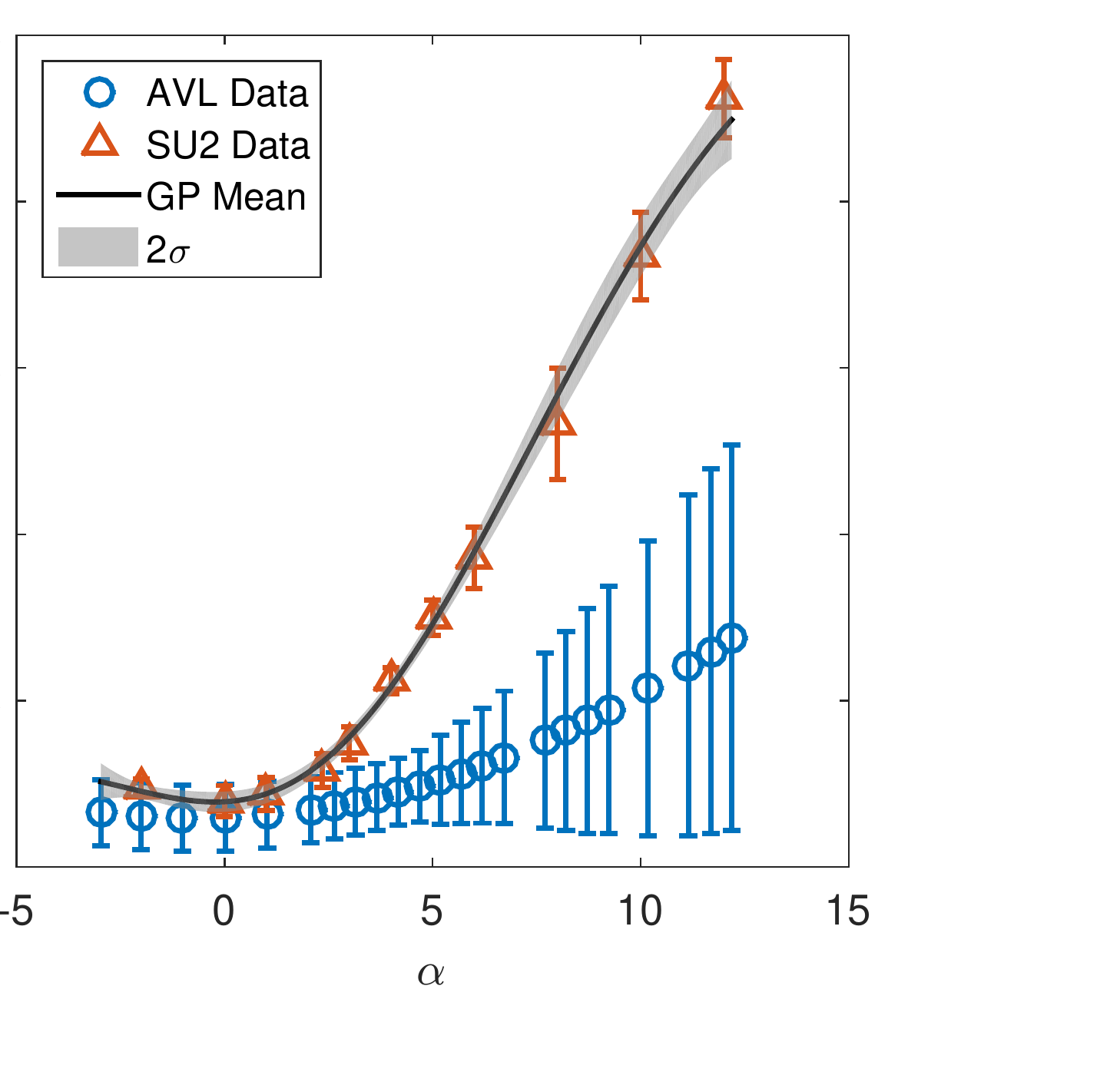} 
    }
    \end{subfigure}
    \hfill
    \begin{subfigure}[3-fidelity fit]{
        \includegraphics[trim=80 180 112 205, clip,             width=.31\textwidth]{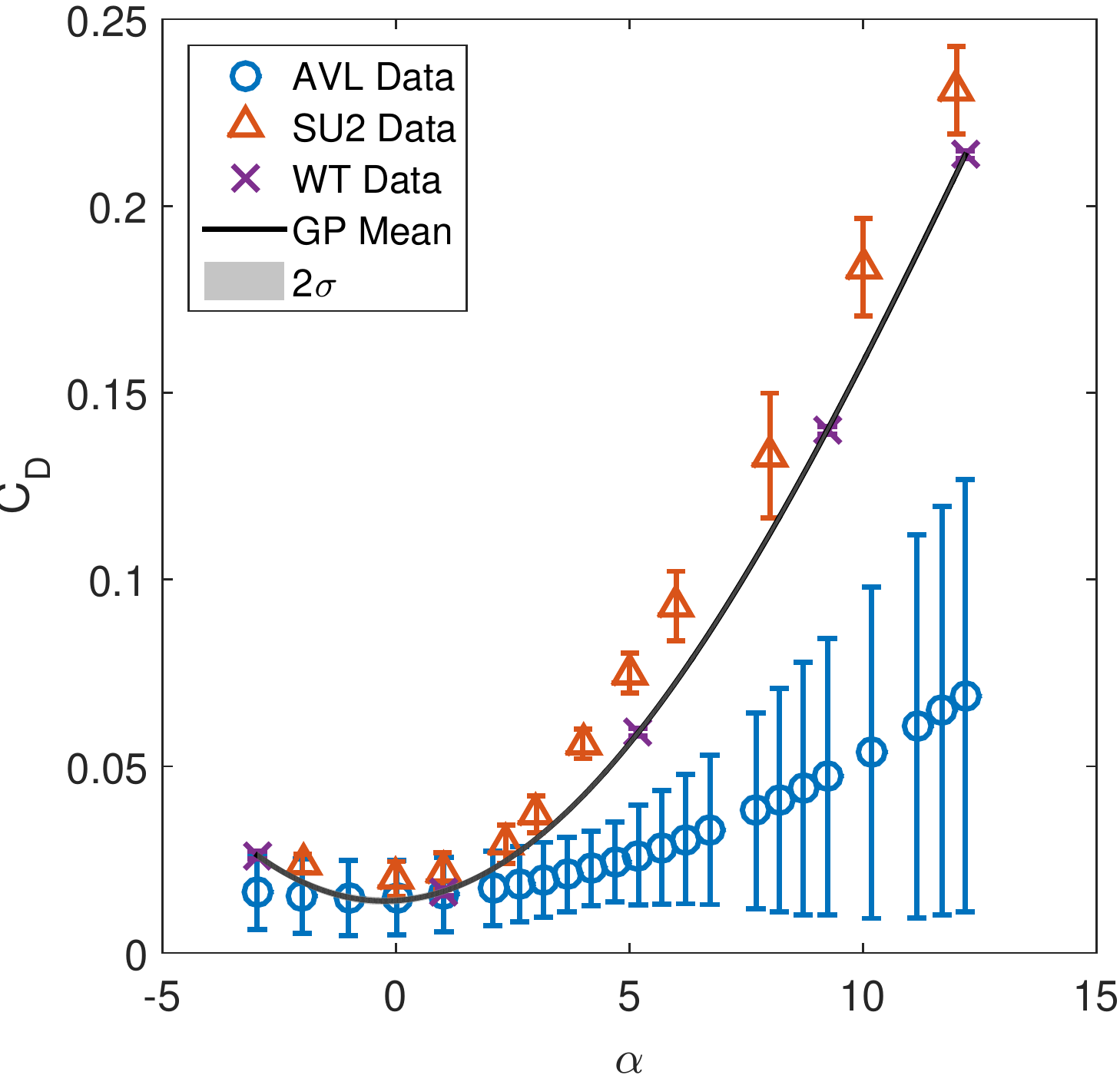} 
    }
    \end{subfigure}
    \caption{$C_D$ vs $\alpha$ for the NASA CRM, using data from multiple sources of varying fidelity.\label{fig:cd_alpha_mf}}
\end{figure}

\begin{figure}
    \centering
    \begin{subfigure}[Single fidelity fit] {
        \includegraphics[trim=80 180 112 205, clip,             width=.31\textwidth]{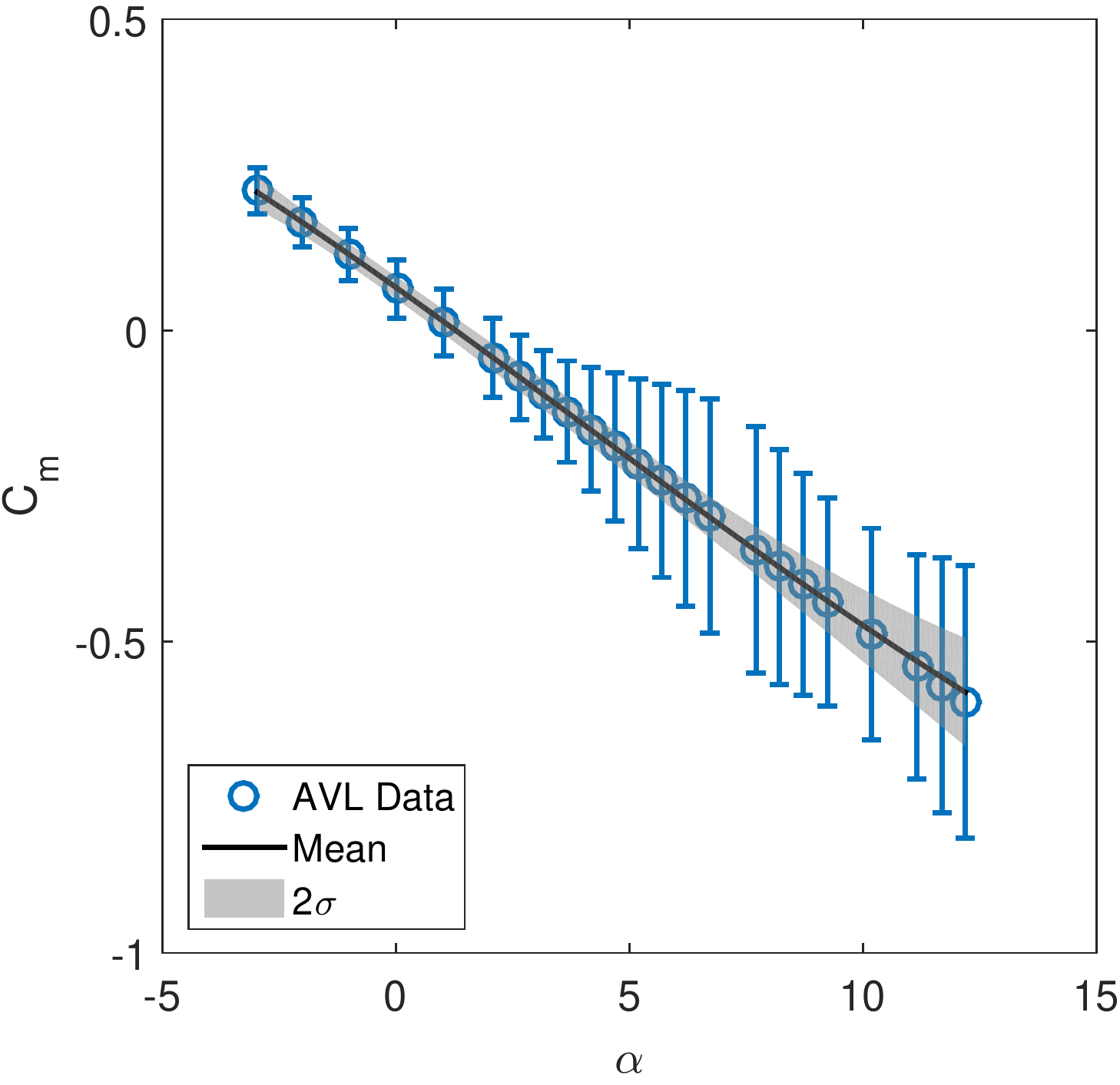} }
    \end{subfigure}
    \hfill
    \begin{subfigure}[2-fidelity fit]{
        \includegraphics[trim=80 180 112 205, clip,             width=.31\textwidth]{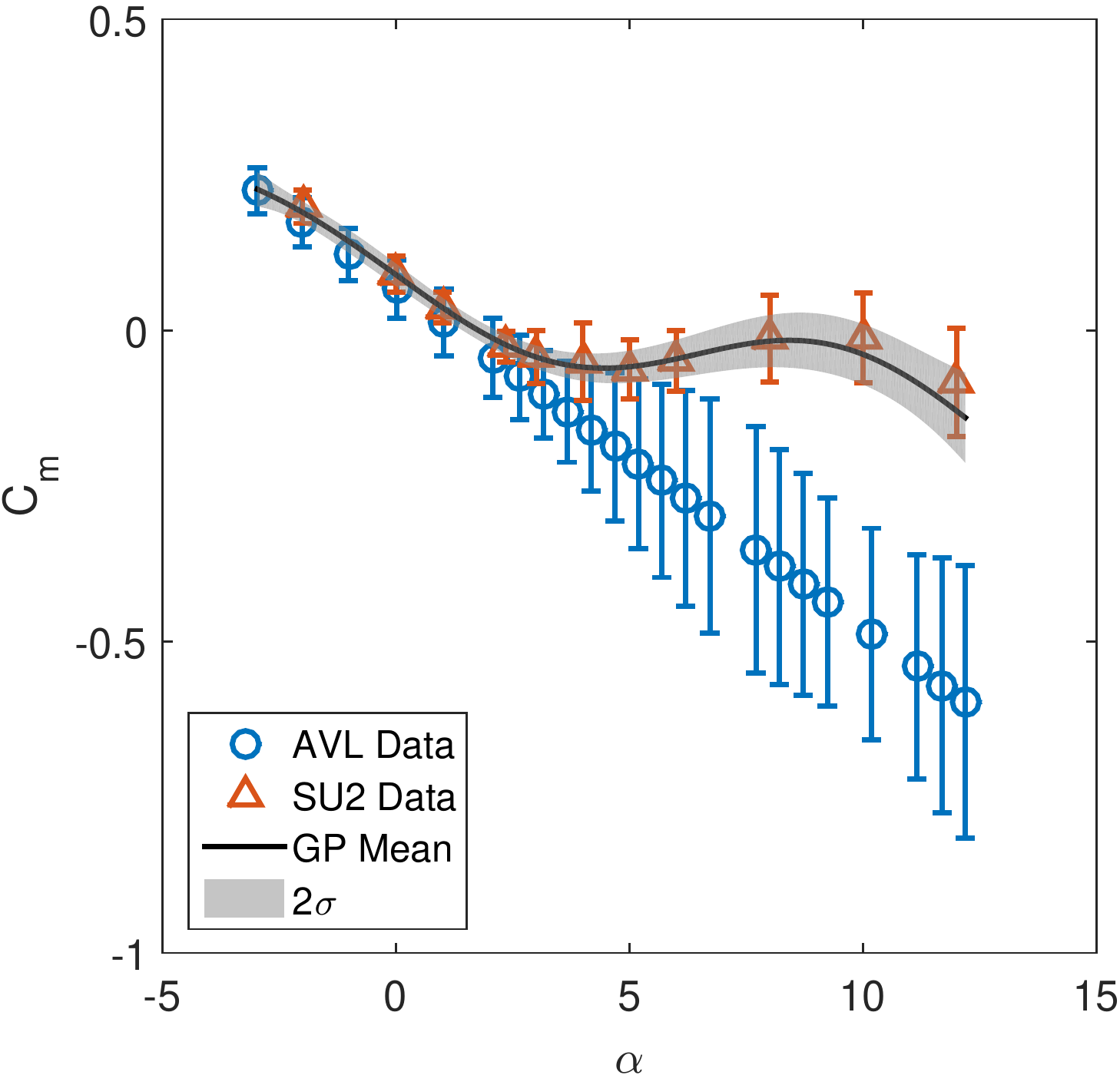} 
    }
    \end{subfigure}
    \hfill
    \begin{subfigure}[3-fidelity fit]{
        \includegraphics[trim=80 180 112 205, clip,             width=.31\textwidth]{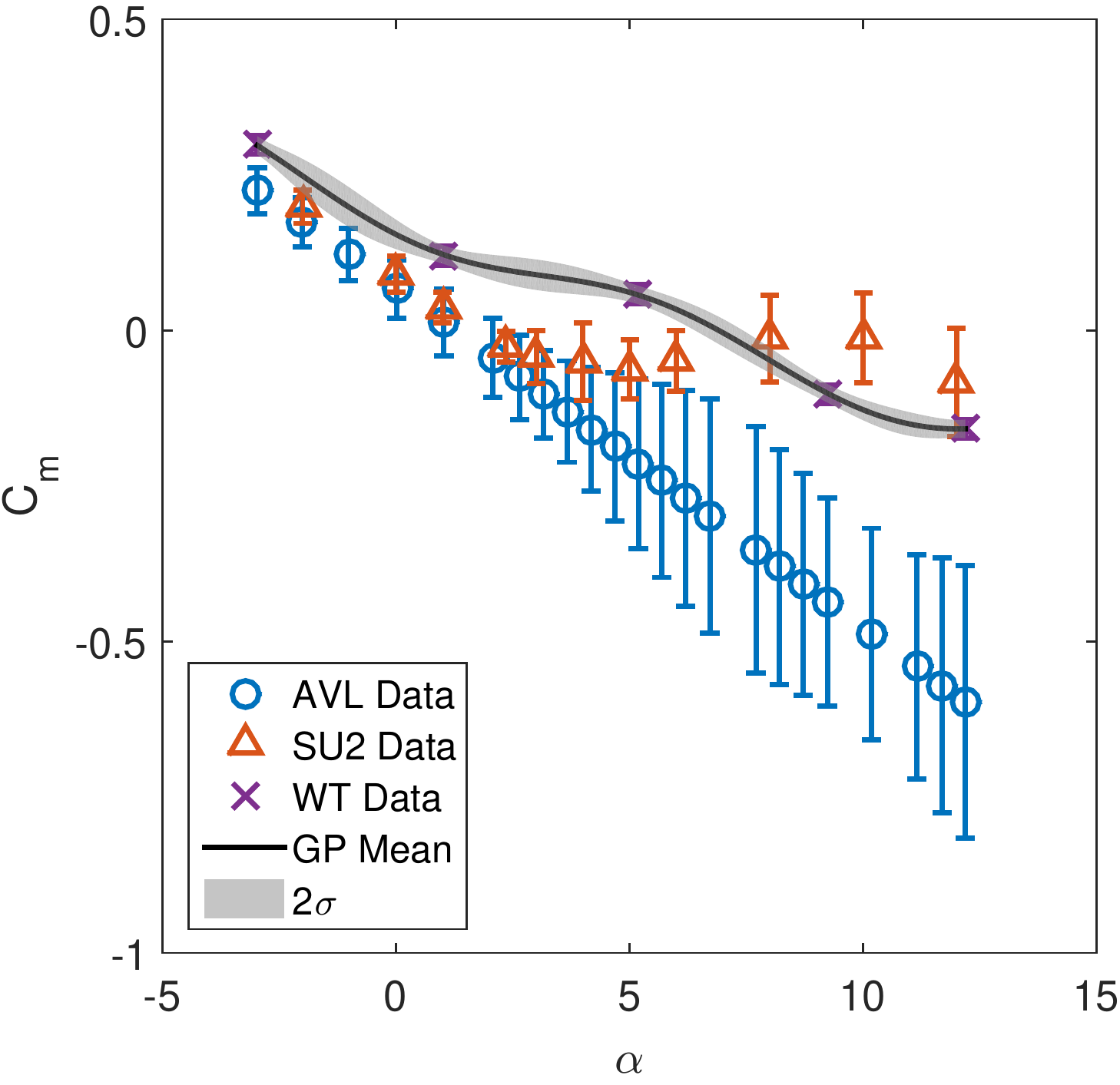} 
    }
    \end{subfigure}
    \caption{$C_m$ vs $\alpha$ for the NASA CRM, using data from multiple sources of varying fidelity.\label{fig:cm_alpha_mf}}
\end{figure}

Using the methodology described in Section \ref{sec:mf_modeling}, $C_L$, $C_D$, and $C_m$ are considered one-dimensional functions of $\alpha$ ($m = 1$) for ease of illustration. It must be mentioned that the methodology is generally applicable to functions of many variables, as is normally the case in aerodynamic databases. %
The results of the multi-fidelity modeling are presented in Figures \ref{fig:cl_alpha_mf} -- \ref{fig:cm_alpha_mf}. For each figure, the solid black line represents the mean predicted by the GP and the grey area represents the $\pm 2\sigma$ interval as predicted by the GP. The left column of figures shows purely the AVL data and a single-fidelity GP fit on that data. In the middle, we introduce the SU2 RANS CFD data with uncertainty bounds informed by the RANS UQ methodology. On the right we introduce a limited set of wind tunnel data points to inform the highest fidelity. For each QoI, the build-up of the database is shown and the distribution of data points across fidelity levels is shown in Table \ref{table:data_points}.

\begin{table}
\centering
    \captionsetup{justification=centering}
    \caption{Number of data points of each fidelity that are used to create Figures \ref{fig:cl_alpha_mf}-\ref{fig:cm_alpha_mf}} 
    \begin{tabular}{|c|c|}
        \hline
        Data Source & Data Points \\ \hline \hline
        Low Fidelity (AVL) & 23 \\ \hline
        Medium Fidelity (CFD) & 11 \\ \hline 
        High Fidelity (Wind Tunnel) & 5 \\ \hline 
    \end{tabular}
    \label{table:data_points}
\end{table}

In Figures \ref{fig:cl_alpha_mf} -- \ref{fig:cm_alpha_mf} the wind tunnel data points are evenly spread across the domain of interest, $-2^\circ < \alpha < 12^\circ$. These data points also have uncertainties associated with them, but these are very small and cannot be seen on this scale in the figures. For the second and third column of figures, the mean and standard deviation predictions are made by the multi-fidelity GP methodology from Section \ref{sec:MF_GP}. 

\begin{figure}
    \centering
    \begin{subfigure}[$C_L$ vs. $\alpha$] {
        \includegraphics[trim=80 180 112 205, clip,             width=.31\textwidth]{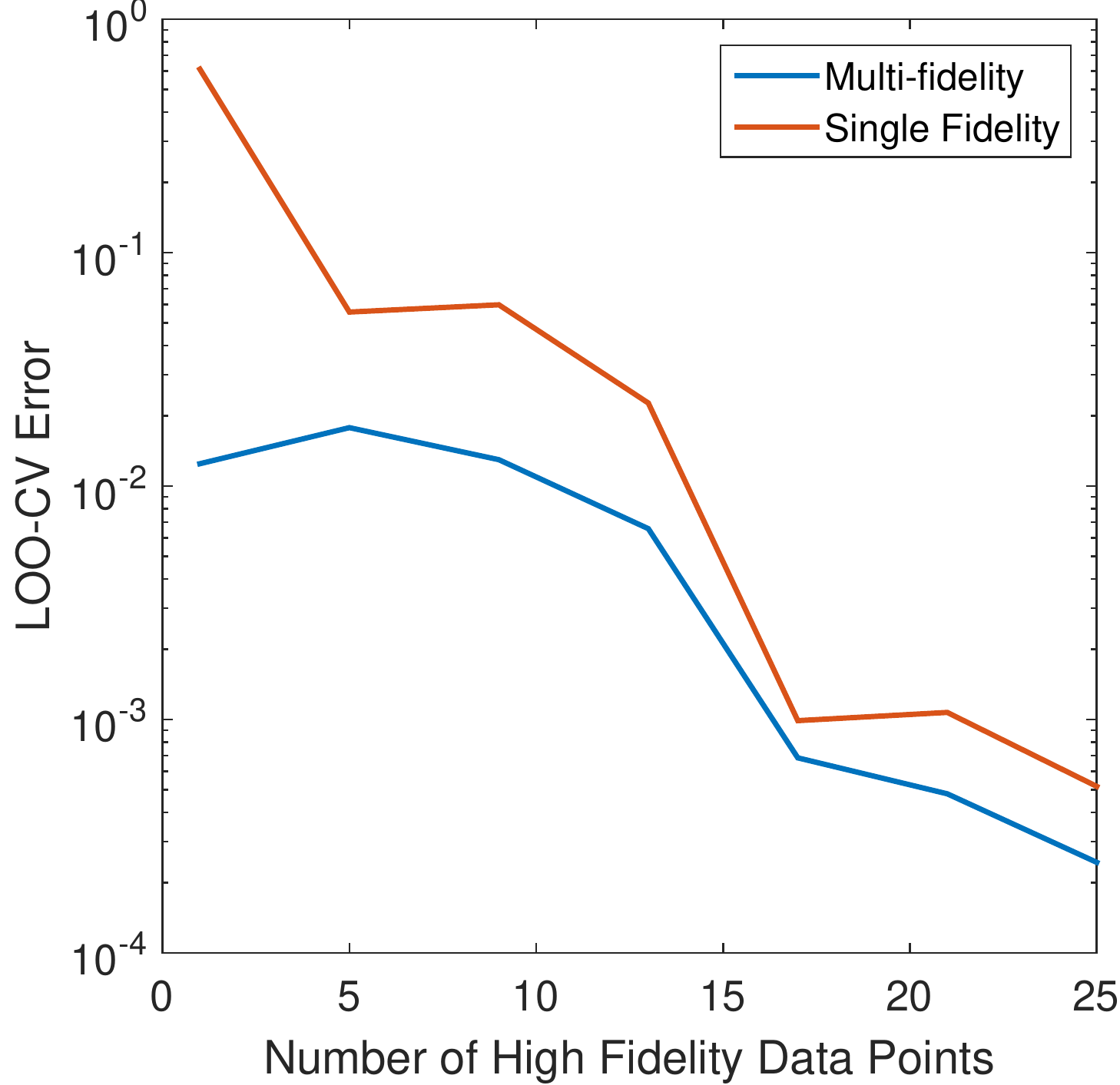} }
    \end{subfigure}
    \hfill
    \begin{subfigure}[$C_D$ vs. $\alpha$]{
        \includegraphics[trim=80 180 112 205, clip,             width=.31\textwidth]{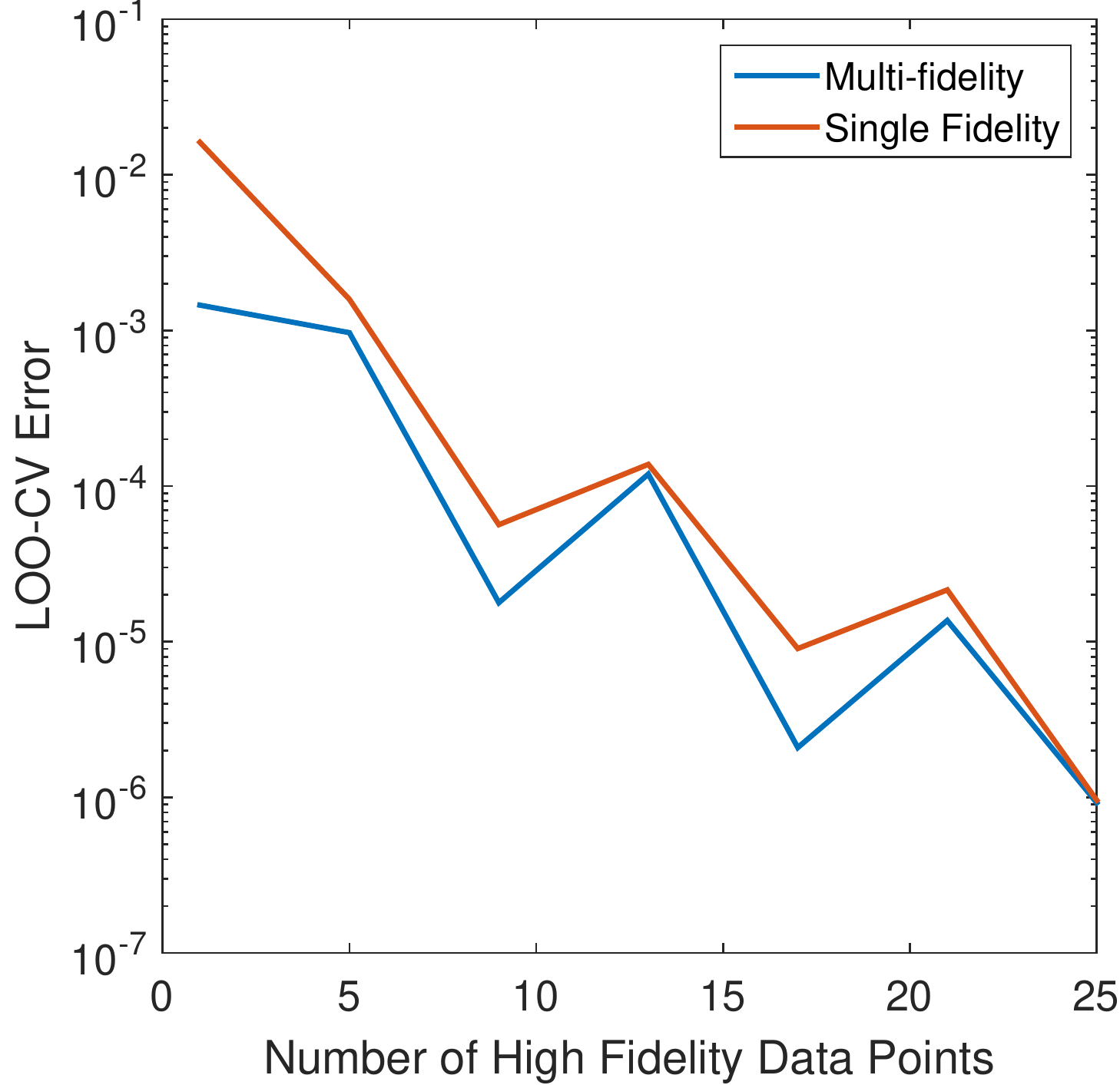} 
    }
    \end{subfigure}
    \hfill
    \begin{subfigure}[$C_m$ vs. $\alpha$]{
        \includegraphics[trim=80 180 112 205, clip,             width=.31\textwidth]{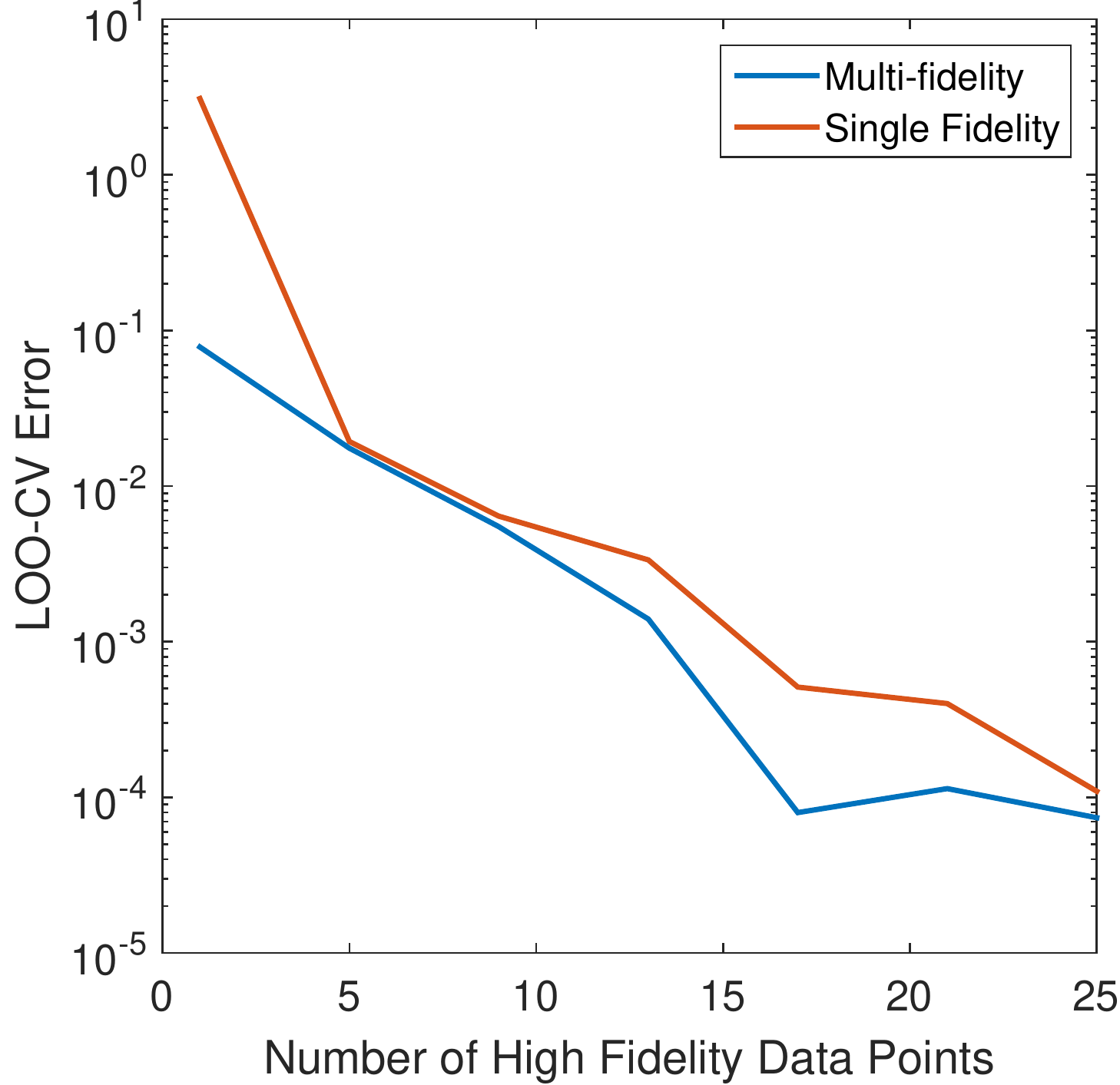} 
    }
    \end{subfigure}
    \caption{Leave-One-Out Cross Validation error when using multi-fidelity data vs. using only high-fidelity data points.\label{fig:mf_vs_hf}}
\end{figure}

The multi-fidelity GPs are able to learn the biases between the different fidelity levels and provide predictions that fit very well with the highest fidelity. To show the benefit of using multi-fidelity data vs. using only high-fidelity data points, the cross-validation error for both cases and for each QoI is presented in Figure \ref{fig:mf_vs_hf}. Due to the small size of the dataset, we use the Leave-One-Out Cross Validation (LOO-CV) error as our performance metric. The multi-fidelity data gives a lower cross-validation error for each QoI. This shows that the lower fidelity information is able to augment the high-fidelity data to improve the accuracy of the predictions involved. This benefit is significant when the high-fidelity data is scarce. As the number of high-fidelity data points increases, the LOO-CV errors converge. In this case, the high-fidelity data being used is evenly spread across the range of angles of attack. %

Another strength of this multi-fidelity GP methodology is apparent when the high-fidelity data is localized to a certain part of the domain. Such a situation might arise if resources are limited and it is not feasible to perform high-fidelity evaluations over the entire domain of interest. It might also be the case that the lower-fidelity simulations are fairly accurate in a certain part of the domain and, consequently, introduce smaller uncertainties in these regions of the domain. For the NASA CRM it is mentioned in Section \ref{sec:crm_uq} that at low angles of attack, where the flow remains attached to the aircraft, RANS CFD simulations are quite successful at predicting performance metrics. This is evidenced by the smaller uncertainty bounds predicted by the RANS UQ methodology in Figure \ref{fig:crm_su2_uq} at $\alpha < 4^\circ$. In this case, an engineer might conclude that highest-fidelity evaluations are not necessary at $\alpha < 4^\circ$ and that sufficient accuracy can be achieved with just the lower-fidelity sources. 

\begin{figure}
    \centering
    \begin{subfigure}[$C_L vs. \alpha$: 3-fidelity fit with localized wind tunnel data] {
        \includegraphics[trim=80 180 112 205, clip,             width=.31\textwidth]{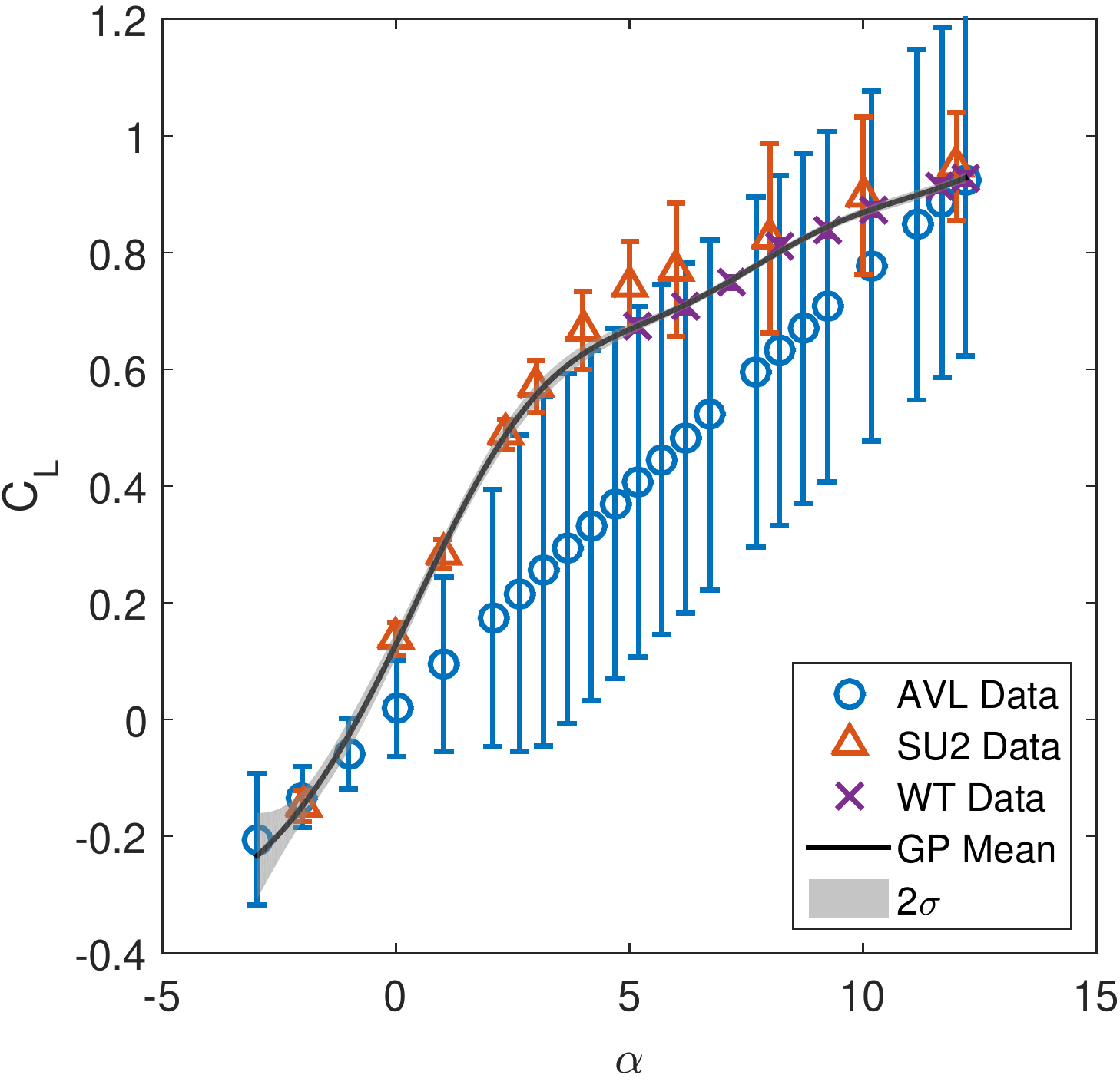} }
    \end{subfigure}
    \hfill
    \begin{subfigure}[$C_L vs. \alpha$: Single fidelity fit with localized wind tunnel data]{
        \includegraphics[trim=80 180 112 205, clip,             width=.31\textwidth]{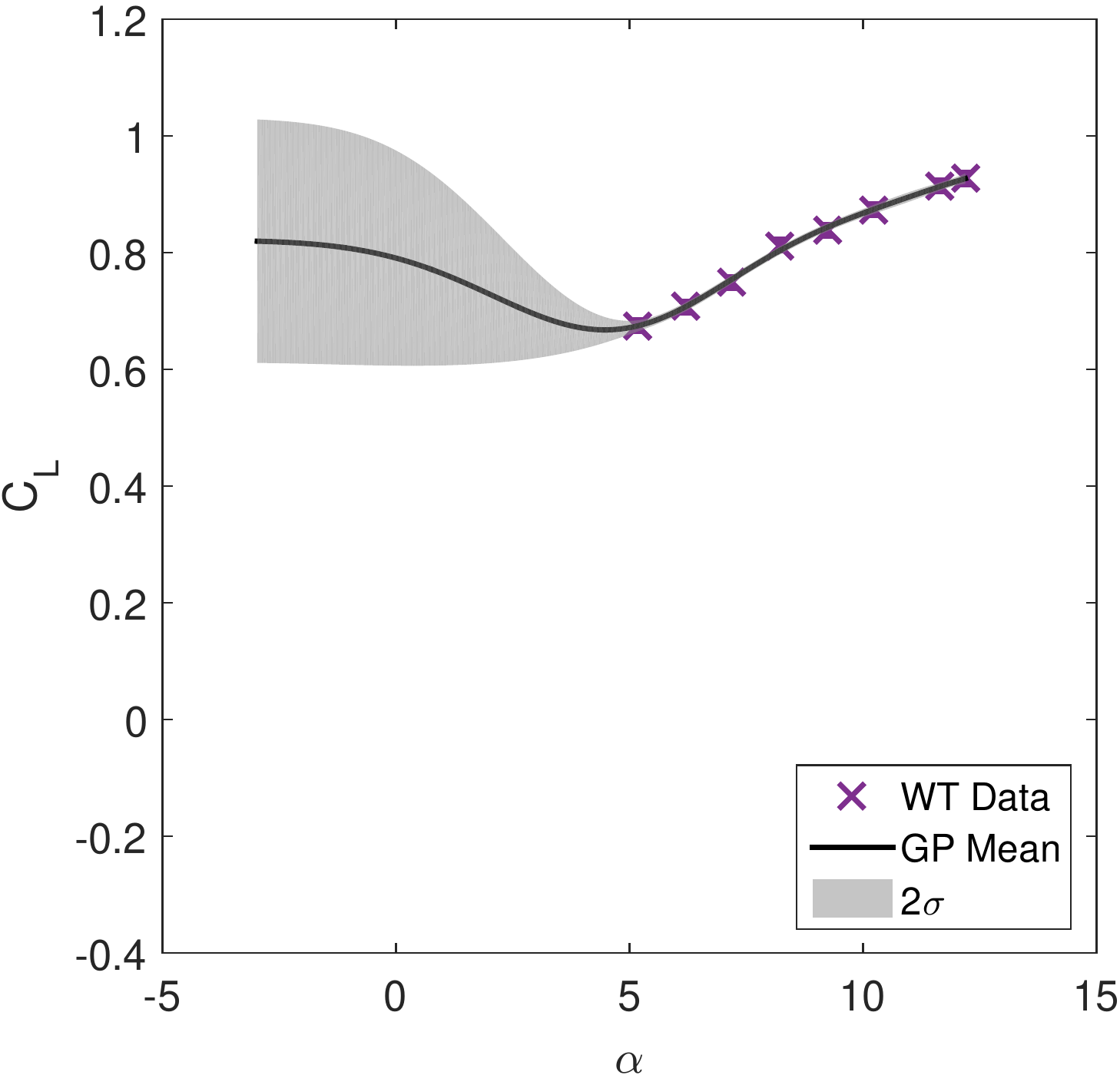} 
    }
    \end{subfigure}
    \hfill
    \begin{subfigure}[$C_L vs. \alpha$: All wind tunnel data to show general trend]{
        \includegraphics[trim=80 180 112 205, clip,             width=.31\textwidth]{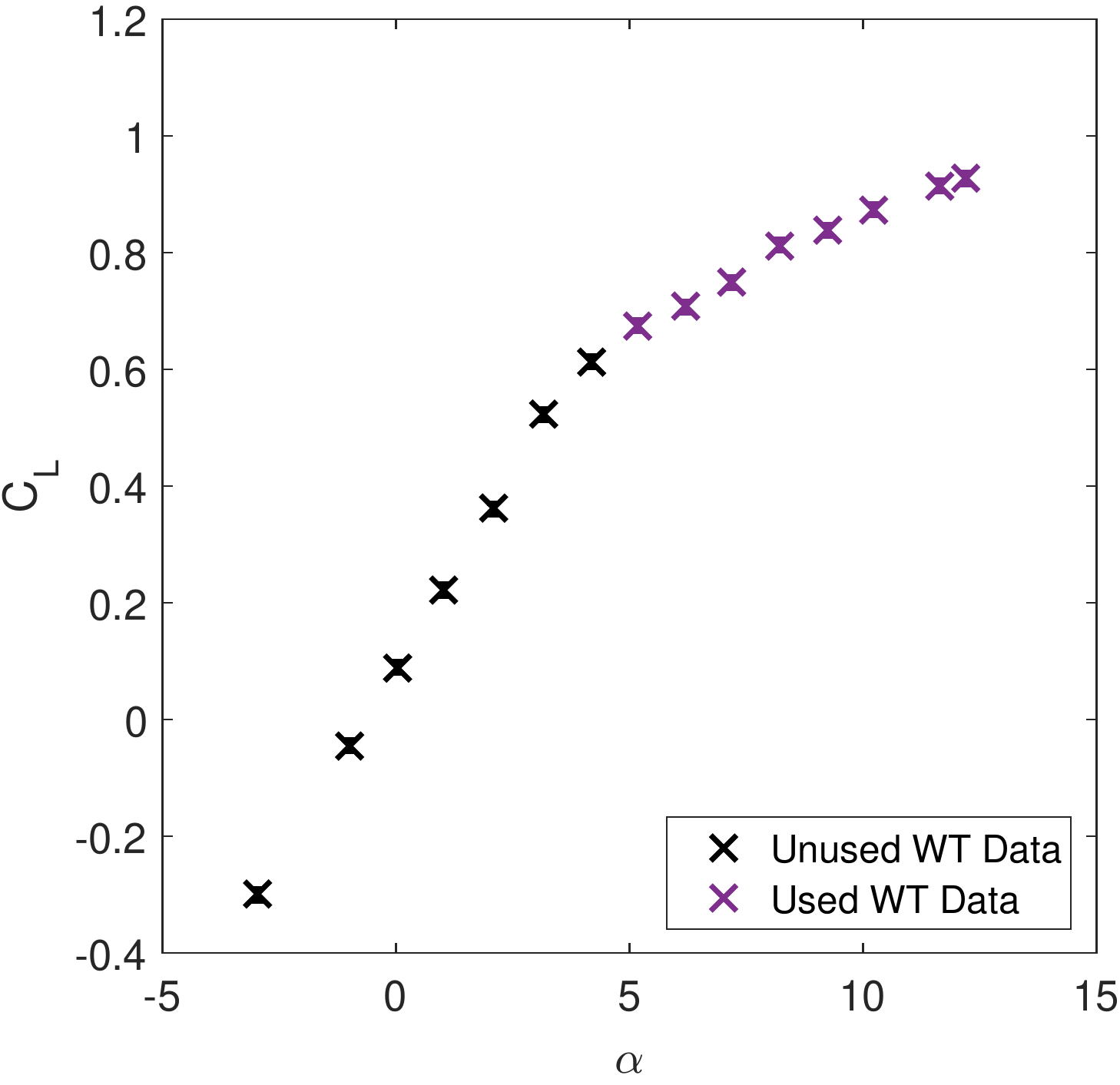} 
    }
    \end{subfigure}

    \begin{subfigure}[$C_D vs. \alpha$: 3-fidelity fit with localized wind tunnel data] {
        \includegraphics[trim=70 180 112 205, clip,             width=.31\textwidth]{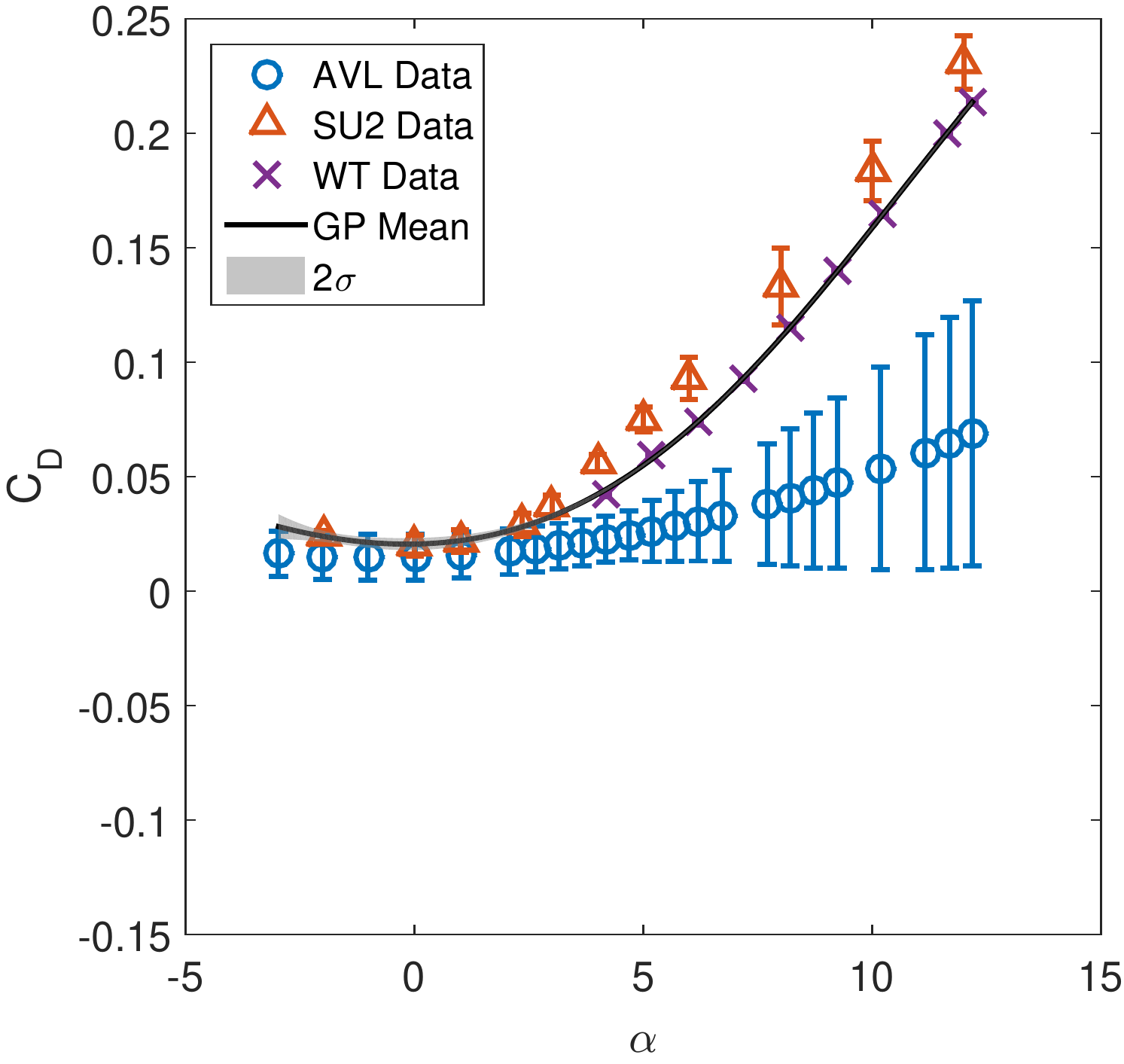} }
    \end{subfigure}
    \hfill
    \begin{subfigure}[$C_D vs. \alpha$: Single fidelity fit with localized wind tunnel data]{
        \includegraphics[trim=70 180 112 205, clip,             width=.31\textwidth]{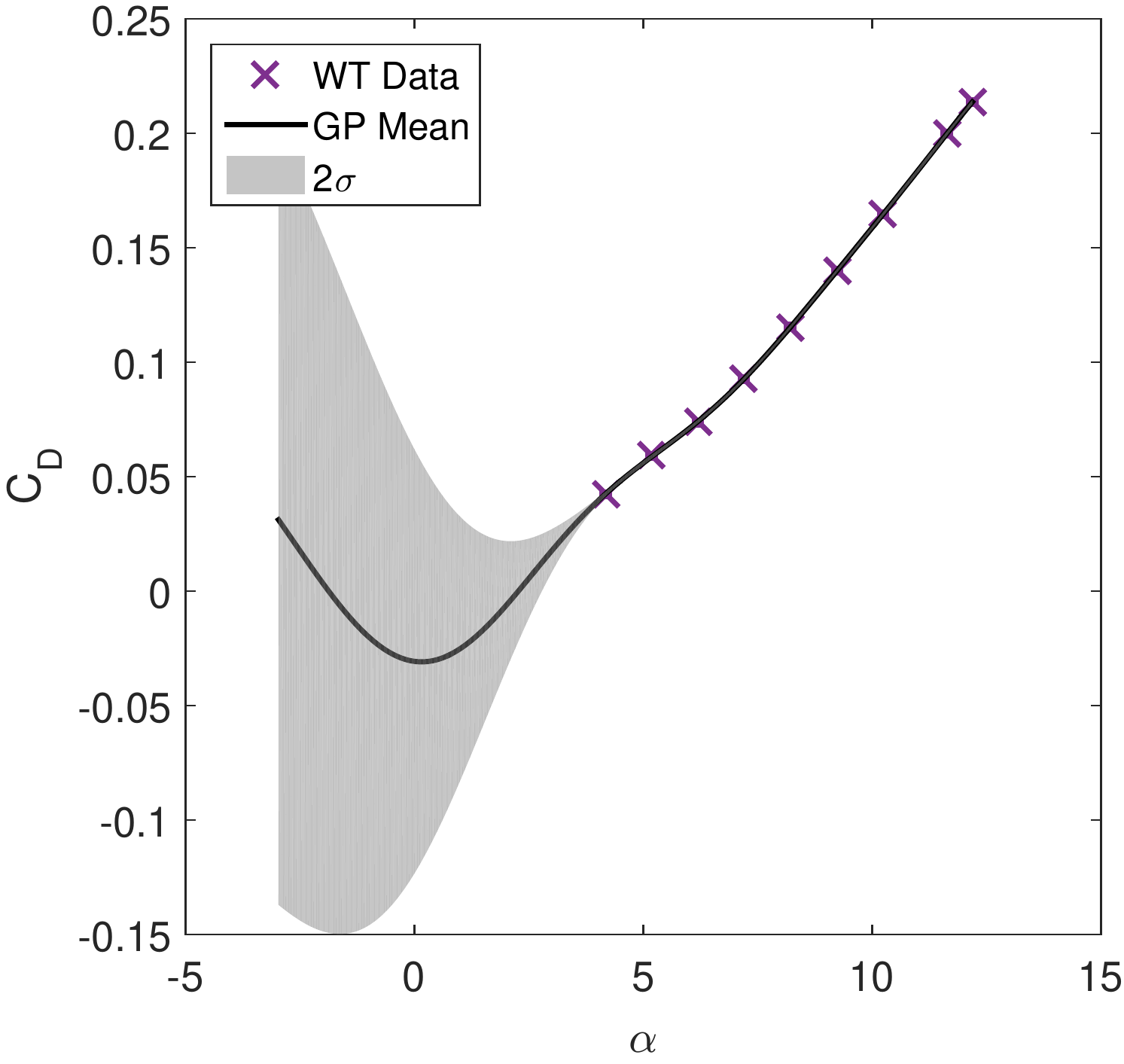} 
    }
    \end{subfigure}
    \hfill
    \begin{subfigure}[$C_D vs. \alpha$: All wind tunnel data to show general trend]{
        \includegraphics[trim=70 180 112 205, clip,             width=.31\textwidth]{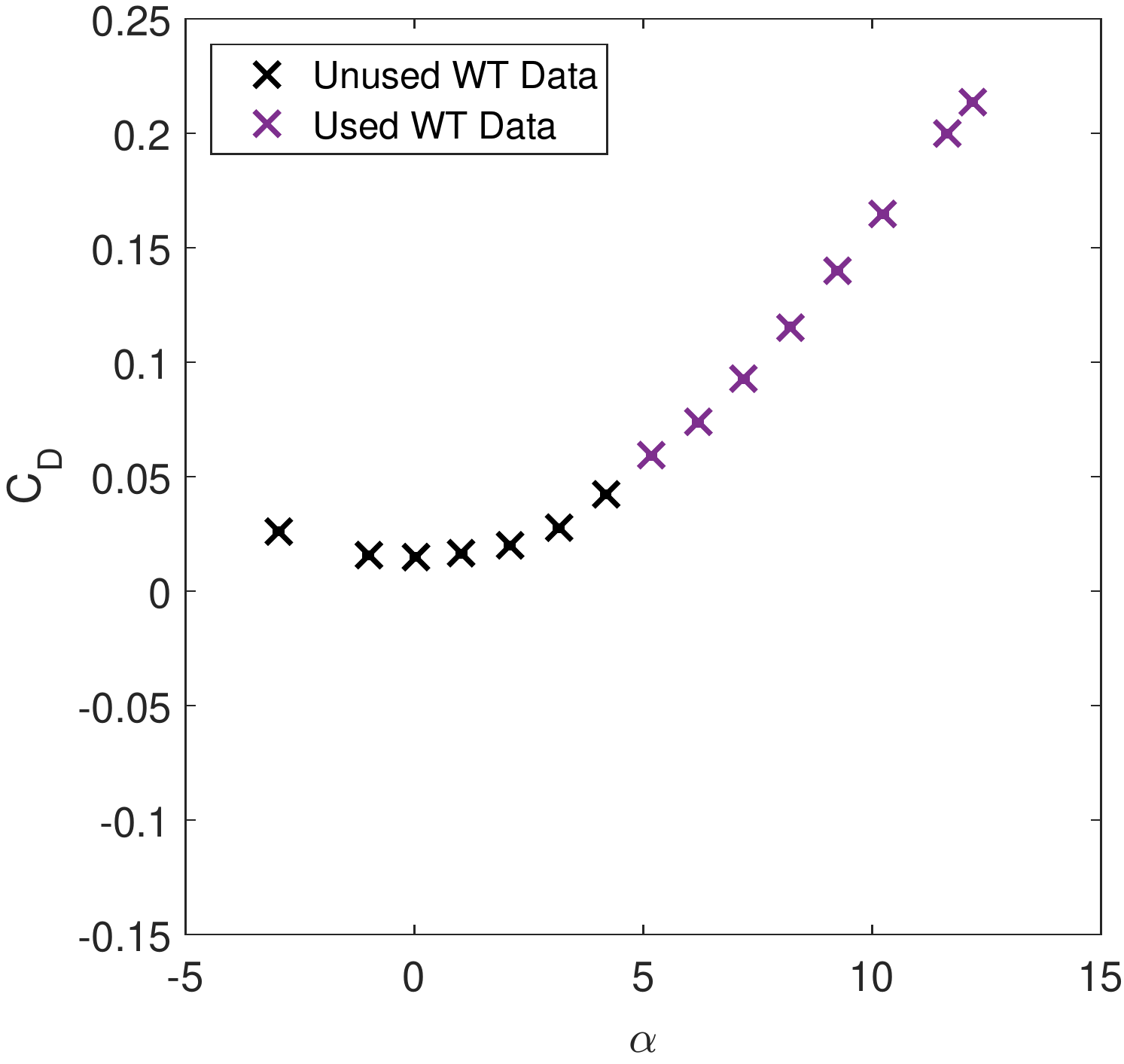} 
    }
    \end{subfigure}

    \begin{subfigure}[$C_m vs. \alpha$: 3-fidelity fit with localized wind tunnel data] {
        \includegraphics[trim=80 180 112 205, clip,             width=.31\textwidth]{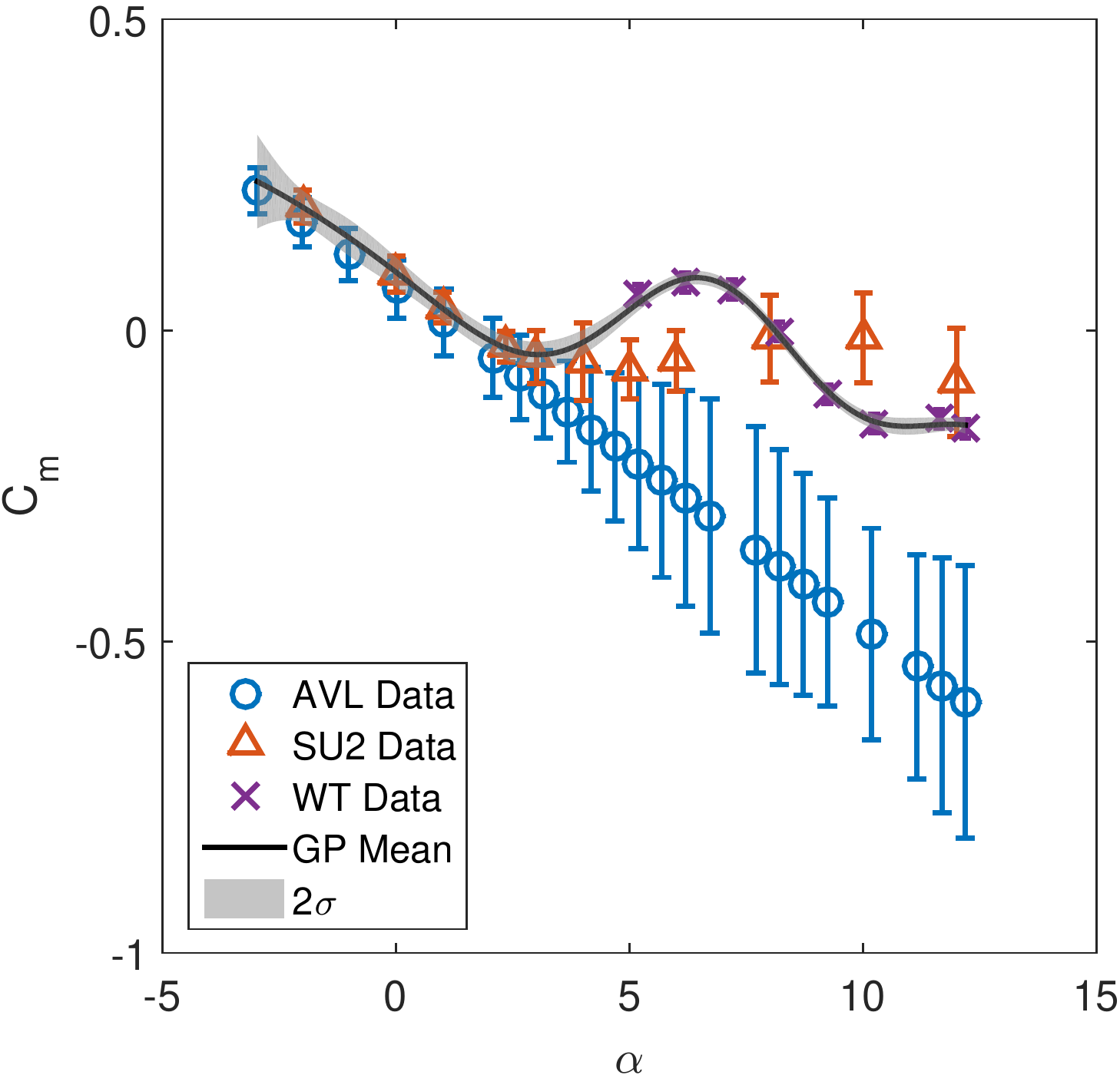} }
    \end{subfigure}
    \hfill
    \begin{subfigure}[$C_m vs. \alpha$: Single fidelity fit with localized wind tunnel data]{
        \includegraphics[trim=80 180 112 205, clip,             width=.31\textwidth]{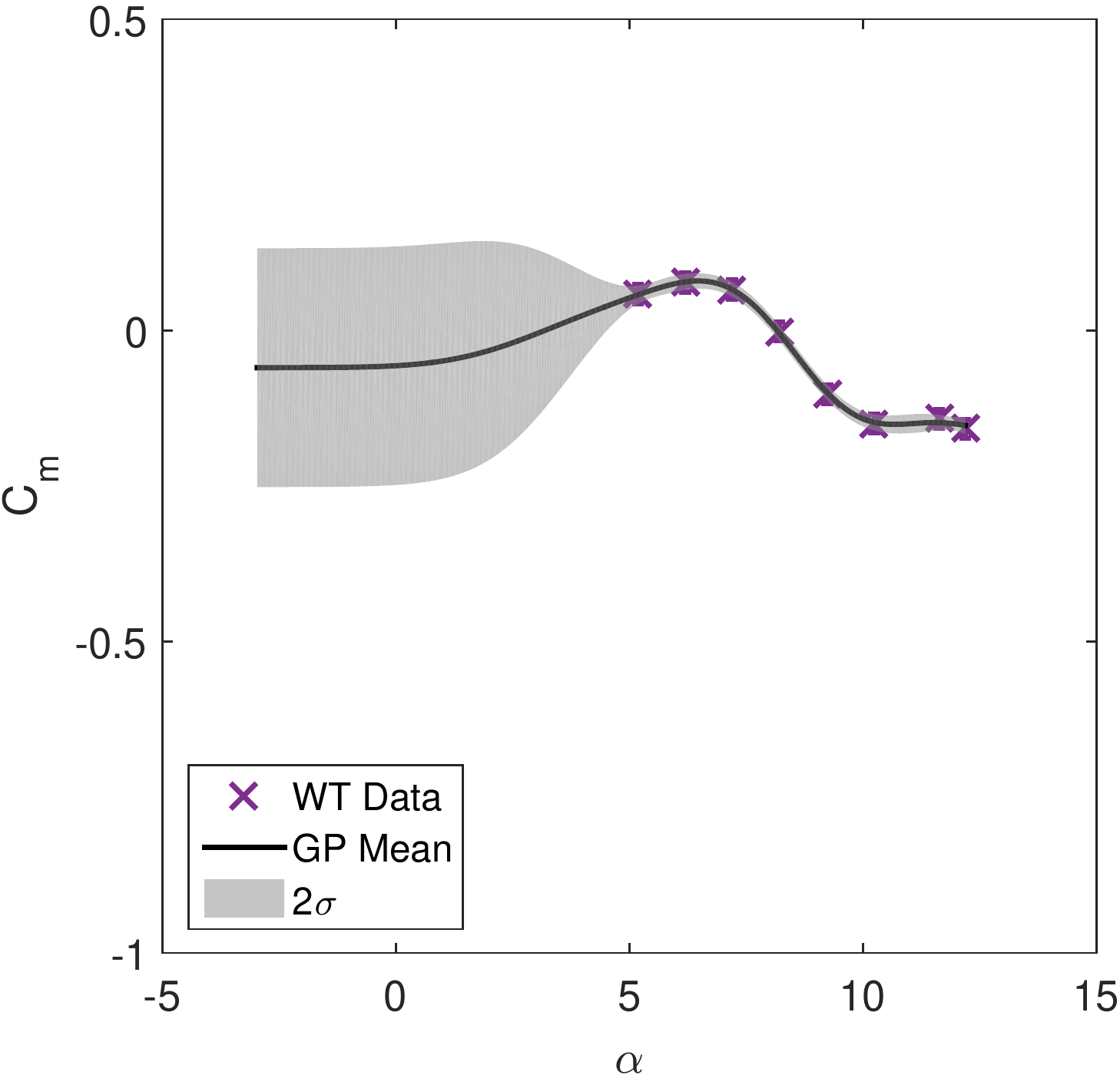} 
    }
    \end{subfigure}
    \hfill
    \begin{subfigure}[$C_m vs. \alpha$: All wind tunnel data to show general trend]{
        \includegraphics[trim=80 180 112 205, clip,             width=.31\textwidth]{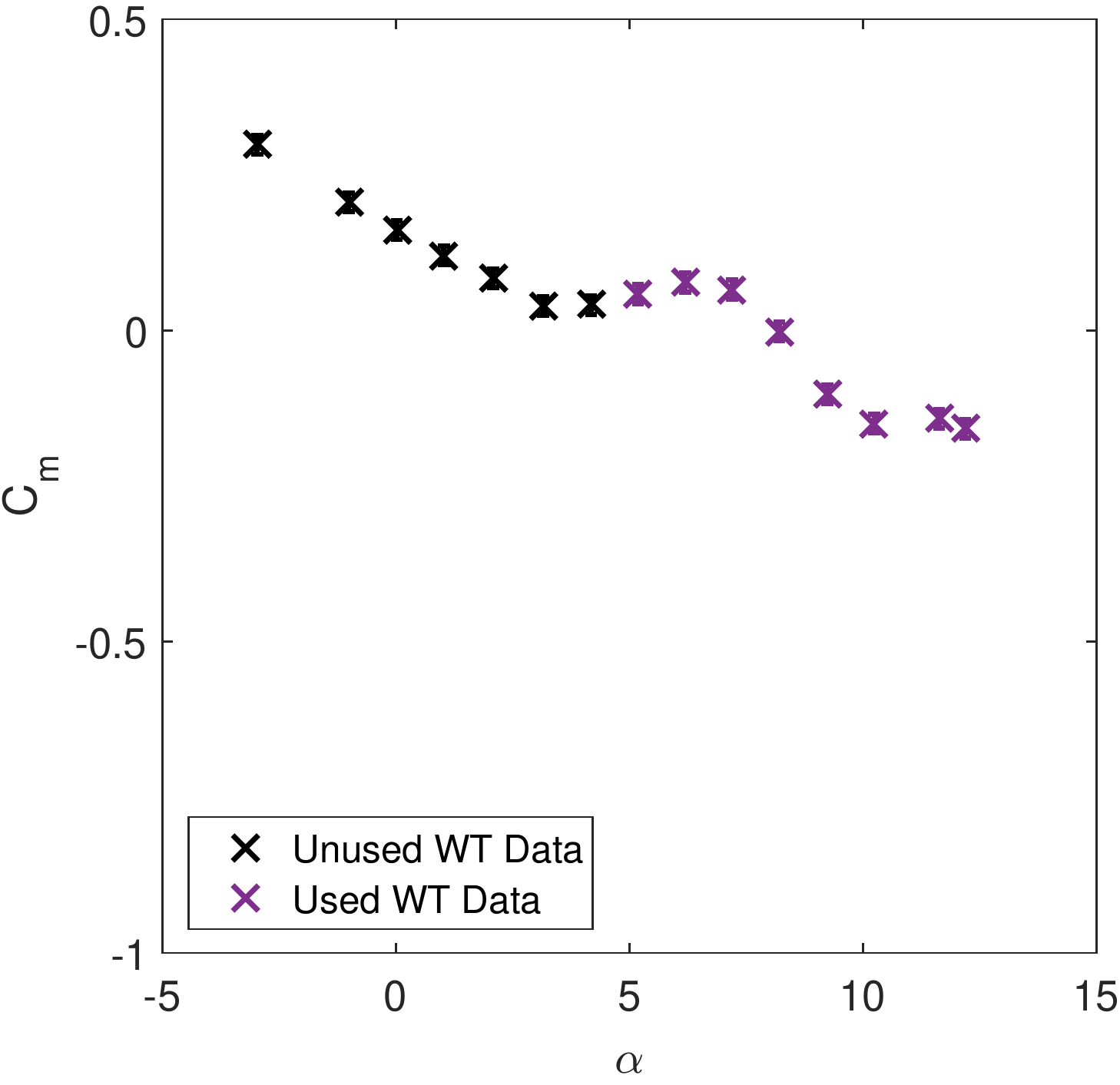} 
    }
    \end{subfigure}
    
    \caption{Showcasing the superior predictive capability of multi-fidelity data fusion when high-fidelity data is localized in the design space $(\alpha \geq 5^\circ)$. The left column represents the multi-fidelity GP formulation result, while the middle column shows the results for the single-fidelity formulation. The right column shows all the wind-tunnel data points to compare the predictive abilities of the two formulations \label{fig:mf_partial}}
    
\end{figure}

To simulate such a situation, a multi-fidelity GP is created that uses AVL and SU2 data that spans the entire domain of interest, but uses wind tunnel evaluations only at high angles of attack $(\alpha > 4^\circ)$. This is a manufactured situation where we choose to ignore some of the wind tunnel data to illustrate the ability of the multi-fidelity GP framework to perform reliably without high-fidelity information that spans the domain of interest. The predictions from the multi-fidelity GP over the domain of interest while using only localized wind tunnel data are presented in the plots in the left column of Figure \ref{fig:mf_partial}. These predictions are compared to the predictions made using a single-fidelity GP trained only on the localized wind tunnel data (shown in the middle column). Lastly, the right column contains the entire wind tunnel data set to show the general trend of the QoI with respect to angle of attack. By comparing the multi-fidelity and single-fidelity GP predictions, it is clear that having accurate low-fidelity data at low angles of attack informs the GP prediction in that region, and allows it to follow the trend of the physical phenomena more accurately than when only the localized high-fidelity data is used.

% --------------------------------------
% Conclusions
% --------------------------------------
\section{Conclusions} \label{sec:conclusions}

In this paper, two main contributions have been presented in the context of multi-fidelity UQ applications of interest to the aerospace engineering industry, and to the simulation-based engineering community: a method to quantify model-form uncertainties in RANS CFD simulations, and a multi-fidelity Gaussian Process framework that combines data sources of varying accuracy to provide better estimates of QoIs and their uncertainties. Both of these methodologies were showcased using a real-world probabilistic aerodynamic database for a full-configuration aircraft, the NASA Common Research Model. 

The RANS UQ methodology uses eigenspace perturbations of the modeled Reynolds' stress tensor to create flow fields that push against the physical realizability constraints of the stress tensor. This methodology provided interval estimates on the QoIs based on the model-form uncertainties associated with turbulence modeling. Simulations at low angles of attack, where the turbulence model is able to accurately capture flow features, had smaller bounds than those performed at high angles of attack, where flow separation is significant and the turbulence model is unable to provide accurate predictions. The predicted bounds did not encapsulate the experimental data due to well known geometric discrepancies between the wind tunnel model and the model used for numerical simulations that resulted from unaccounted aeroelastic twist.

Previous work in multi-fidelity Gaussian processes was advanced by introducing noise in the observations of the QoIs. The RANS CFD simulations and associated uncertainty predictions were combined with low-fidelity data from AVL simulations, and high-fidelity data from wind tunnel experiments to create multi-fidelity surrogate models for $C_L$, $C_D$, and $C_m$ for the NASA CRM. These multi-fidelity fits were more accurate than single-fidelity GP fits on just the high-fidelity data at a significantly lower cost, showing that lower-fidelity data that is well correlated with higher-fidelity data can serve to augment high-fidelity data to improve predictive capabilities.  This is especially true in multi-dimensional operating spaces where the quantity of interest would require the use of a prohibitive number of high-fidelity simulations or wind-tunnel data points.

Finally, we would like the current work to serve as an early (and, necessarily, incomplete) example of the direction of evolution that would be beneficial for future engineering simulations. As industry increases its reliance on numerical simulations for design and analysis, it is essential to understand and address the shortcomings of such simulations. Quantifying uncertainties introduced by these predictive methodologies is the first step in extending their value. It enables the use of a reliability-based design process instead of the traditional, deterministic design methodology that makes no allowance for modeling uncertainties. Additionally, combining predictions from all the analysis methods that are used in the design process improves their predictive capability with lower computational cost. 

%

% --------------------------------------
% Acknowledgements
% --------------------------------------
\acknowledgements
The authors would like to thank Andrew Cary from The Boeing Company for multiple conversations on the material and The Boeing Company for funding this research under grant number 131303 [IC2017-0559].

%% appendix sections are then done as normal sections and after Acknowledgements



%\begin{thebibliography}{-8}


%\small{
%\bibitem{key}

%...

%}

%\end{thebibliography}


% --------------------------------------
% Biblio
% --------------------------------------

\bibliographystyle{IJ4UQ_Bibliography_Style}

\bibliography{References}

\begin{thebibliography}{10}

\bibitem{queipo2005surrogate}
Queipo, N.V., Haftka, R.T., Shyy, W., Goel, T., Vaidyanathan, R., and Tucker,
  P.K., Surrogate-based analysis and optimization, {\em Progress in aerospace
  sciences}, 41(1):1--28, 2005.

\bibitem{gorissen2010surrogate}
Gorissen, D., Couckuyt, I., Demeester, P., Dhaene, T., and Crombecq, K., A
  surrogate modeling and adaptive sampling toolbox for computer based design,
  {\em Journal of Machine Learning Research}, 11(Jul):2051--2055, 2010.

\bibitem{jeong2005efficient}
Jeong, S., Murayama, M., and Yamamoto, K., Efficient optimization design method
  using kriging model, {\em Journal of aircraft}, 42(2):413--420, 2005.

\bibitem{peherstorfer_survey_2018}
Peherstorfer, B., Willcox, K., and Gunzburger, M., Survey of {Multifidelity}
  {Methods} in {Uncertainty} {Propagation}, {Inference}, and {Optimization},
  {\em SIAM Review}, 60(3):550--591, January 2018.

\bibitem{forrester_multi-fidelity_2007}
Forrester, A.I., Sóbester, A., and Keane, A.J., Multi-fidelity optimization
  via surrogate modelling, {\em Proceedings of the Royal Society A:
  Mathematical, Physical and Engineering Sciences}, 463(2088):3251--3269,
  December 2007.

\bibitem{reliability}
Harr, M.E., {\em Reliability based design}, Dover, New York, 1996.

\bibitem{wendorff_combining_2016}
Wendorff, A., Combining {Uncertainty} and {Sensitivity} {Using}
  {Multi}-{Fidelity} {Probabilistic} {Aerodynamic} {Databases} for {Aircraft}
  {Maneuvers}, PhD thesis, Stanford, December 2016.

\bibitem{ng_multifidelity_2014}
Ng, L.W.T. and Willcox, K.E., Multifidelity approaches for optimization under
  uncertainty, {\em International Journal for Numerical Methods in
  Engineering}, 100(10):746--772, December 2014.

\bibitem{multif}
Fenrich, R.W., Menier, V., Avery, P., and Alonso, J.J.
\newblock Reliability-based design optimization of a supersonic nozzle.
\newblock In {\em Proceedings of the 7th European Conference on Computational
  Fluid Dynamics}. ECCOMAS, 2018.

\bibitem{kennedy_predicting_2000}
Kennedy, M. and O'Hagan, A., Predicting the output from a complex computer code
  when fast approximations are available, {\em Biometrika}, 87(1):1--13, March
  2000.

\bibitem{le_gratiet_recursive_2014}
Gratiet, L.L. and Garnier, J., Recursive {Co}-{Kriging} {Model} for {Design} of
  {Computer} {Experiments} with {Multiple} {Levels} of {Fidelity}, {\em
  International Journal for Uncertainty Quantification}, 4(5):365--386, 2014.

\bibitem{huang_sequential_2006}
Huang, D., Allen, T.T., Notz, W.I., and Miller, R.A., Sequential kriging
  optimization using multiple-fidelity evaluations, {\em Structural and
  Multidisciplinary Optimization}, 32(5):369--382, September 2006.

\bibitem{robinson2008surrogate}
Robinson, T., Eldred, M., Willcox, K., and Haimes, R., Surrogate-based
  optimization using multifidelity models with variable parameterization and
  corrected space mapping, {\em Aiaa Journal}, 46(11):2814--2822, 2008.

\bibitem{rasmussen_gaussian_2006}
Rasmussen, C.E. and Williams, C.K.I., {\em Gaussian processes for machine
  learning}, Adaptive computation and machine learning, MIT Press, Cambridge,
  Mass, 2006.
\newblock OCLC: ocm61285753.

\bibitem{gratiet_multi-fidelity_nodate}
Gratiet, L.L., Multi-fidelity {Gaussian} process regression for computer
  experiments, PhD thesis, Université Paris-Diderot, 2013.

\bibitem{pope_2000}
Pope, S.B., {\em Turbulent Flows}, Cambridge University Press, 2000.

\bibitem{iaccarino_eig_pert}
Iaccarino, G., Mishra, A., and Ghili, S., Eigenspace perturbations for
  uncertainty estimation of single-point turbulence closures, {\em Physical
  Review Fluids}, 2(2), 2017.

\bibitem{mishra_uncertainty_2019}
Mishra, A.A., Mukhopadhaya, J., Iaccarino, G., and Alonso, J., Uncertainty
  {Estimation} {Module} for {Turbulence} {Model} {Predictions} in {SU}2, {\em
  AIAA Journal}, 57(3):1066--1077, March 2019.

\bibitem{Gerolymos2016AlgebraicPA}
Gerolymos, G.A. and Vallet, I., Algebraic proof and application of lumley's
  realizability triangle, 2016.

\bibitem{schumann1977realizability}
Schumann, U., Realizability of reynolds-stress turbulence models, {\em The
  Physics of Fluids}, 20(5):721--725, 1977.

\bibitem{speziale1994realizability}
Speziale, C.G., Abid, R., and Durbin, P.A., On the realizability of reynolds
  stress turbulence closures, {\em Journal of Scientific Computing},
  9(4):369--403, 1994.

\bibitem{2014realizability}
Mishra, A.A. and Girimaji, S.S., On the realizability of pressure--strain
  closures, {\em Journal of Fluid Mechanics}, 755:535--560, 2014.

\bibitem{banerjee2007presentation}
Banerjee, S., Krahl, R., Durst, F., and Zenger, C., Presentation of anisotropy
  properties of turbulence, invariants versus eigenvalue approaches, {\em
  Journal of Turbulence}, 8:N32, 2007.

\bibitem{mishra_perturbations_2019}
Mishra, A.A. and Iaccarino, G., Theoretical analysis of tensor perturbations
  for uncertainty quantification of reynolds averaged and subgrid scale
  closures, {\em Physics of Fluids}, 31(7):075101, 2019.

\bibitem{emory2011characterizing}
Emory, M., Terrapon, V., Pecnik, R., and Iaccarino, G., Characterizing the
  operability limits of the hyshot ii scramjet through rans simulations, In
  {\em 17th {AIAA} international space planes and hypersonic systems and
  technologies conference}, p. 2282, 2011.

\bibitem{aiaajets}
Mishra, A.A. and Iaccarino, G., Uncertainty estimation for reynolds-averaged
  navier--stokes predictions of high-speed aircraft nozzle jets, {\em AIAA
  Journal}, pp. 3999--4004, 2017.

\bibitem{envelopingmodels}
Mishra, A. and Iaccarino, G., Rans predictions for high-speed flows using
  enveloping models, {\em arXiv preprint arXiv:1704.01699}, 2017.

\bibitem{emory2016uncertainty}
Emory, M., Iaccarino, G., and Laskowski, G.M., Uncertainty quantification in
  turbomachinery simulations, In {\em ASME Turbo Expo 2016: Turbomachinery
  Technical Conference and Exposition}, pp. V02CT39A028--V02CT39A028. American
  Society of Mechanical Engineers, 2016.

\bibitem{razaaly2019optimization}
Razaaly, N., Gori, G., Iaccarino, G., and Congedo, P., Optimization of an orc
  supersonic nozzle under epistemic uncertainties due to turbulence models, In
  {\em GPPS 2019}, 2019.

\bibitem{garcia2014quantifying}
Garc{\'\i}a-S{\'a}nchez, C., Philips, D., and Gorl{\'e}, C., Quantifying inflow
  uncertainties for cfd simulations of the flow in downtown oklahoma city, {\em
  Building and Environment}, 78:118--129, 2014.

\bibitem{ricci2015local}
Ricci, A., Kalkman, I., Blocken, B., Repetto, M., Burlando, M., and Freda, A.,
  Local-scale forcing effects on wind flows in an urban environment, In {\em
  Proceedings, International Workshop on Physical Modelling of Flow and
  Dispersion Phenomena PHYSMOD2015}, pp. 7--9, 2015.

\bibitem{gorle2013framework}
Gorl{\'e}, C. and Iaccarino, G., A framework for epistemic uncertainty
  quantification of turbulent scalar flux models for reynolds-averaged
  navier-stokes simulations, {\em Physics of Fluids}, 25(5):055105, 2013.

\bibitem{rivers_further_2012}
Rivers, M., Hunter, C., and Campbell, R., Further {Investigation} of the
  {Support} {System} {Effects} and {Wing} {Twist} on the {NASA} {Common}
  {Research} {Model}, In {\em 30th {AIAA} {Applied} {Aerodynamics}
  {Conference}}, New Orleans, Louisiana, June 2012. American Institute of
  Aeronautics and Astronautics.

\bibitem{rivers_experimental_2010}
Rivers, M. and Dittberner, A., Experimental {Investigations} of the {NASA}
  {Common} {Research} {Model} ({Invited}), In {\em 28th {AIAA} {Applied}
  {Aerodynamics} {Conference}}, Chicago, Illinois, June 2010. American
  Institute of Aeronautics and Astronautics.

\bibitem{morrison20094th}
Morrison, J.H.
\newblock 4th {AIAA} {CFD} drag prediction workshop, 2009.

\bibitem{levy2013summary}
Levy, D., Laflin, K., Vassberg, J., Tinoco, E., Mani, M., Rider, B., Brodersen,
  O., Crippa, S., Rumsey, C., Wahls, R., , Summary of data from the fifth aiaa
  cfd drag prediction workshop, In {\em 51st AIAA Aerospace Sciences Meeting
  including the New Horizons Forum and Aerospace Exposition}, p.~46, 2013.

\bibitem{morrison20166th}
Morrison, J.
\newblock 6th {AIAA} {CFD} drag prediction workshop.[online] aiaa, 2016.

\bibitem{roy2017summary}
Roy, C.J., Summary of data from the sixth {AIAA} {CFD} drag prediction
  workshop: case 1 code verification, In {\em 55th AIAA Aerospace Sciences
  Meeting}, p. 1206, 2017.

\bibitem{tinoco2017summary}
Tinoco, E.N., Brodersen, O., Keye, S., and Laflin, K., Summary of data from the
  sixth {AIAA} {CFD} drag prediction workshop: {CRM} cases 2 to 5, In {\em 55th
  AIAA Aerospace Sciences Meeting}, p. 1208, 2017.

\bibitem{su2_aiaajournal}
Economon, T.D., Palacios, F., Copeland, S.R., Lukaczyk, T.W., and Alonso, J.J.,
  Su2: An open-source suite for multiphysics simulation and design, {\em AIAA
  Journal}, 54(3):828--846, 2016.

\bibitem{menter1994two}
Menter, F.R., Two-equation eddy-viscosity turbulence models for engineering
  applications, {\em AIAA journal}, 32(8):1598--1605, 1994.

\bibitem{menter2003ten}
Menter, F.R., Kuntz, M., and Langtry, R., Ten years of industrial experience
  with the sst turbulence model, {\em Turbulence, heat and mass transfer},
  4(1):625--632, 2003.

\bibitem{drela2008athena}
Drela, M. and Youngren, H., Athena vortex lattice ({AVL}), {\em Computer
  software. AVL}, 4, 2008.

\end{thebibliography}
\end{document}